\documentclass[LectureNotes]{SciPost}
\pdfoutput=1

\usepackage{float}
\restylefloat{table}
\usepackage{microtype}
\usepackage{xcolor}
\usepackage[most]{tcolorbox}
\usepackage[utf8]{inputenc}
\usepackage{amsmath,amssymb,amsfonts,amsthm,mathtools,mathrsfs,euscript}
\usepackage{babel}
\usepackage{slashed}
\usepackage[sort&compress,numbers]{natbib}
\usepackage{tocloft}
\usepackage{svg}
\usepackage{tikz}
\usepackage{tikz-feynman}
\usepackage{feynmp-auto}
\usepackage{fontenc}

\definecolor{yellow1}{HTML}{ffffcc}
\definecolor{myviolet}{HTML}{007B8B}
\definecolor{mylavender}{HTML}{DEFEFF}
\definecolor{soft}{HTML}{fff6ed}
\definecolor{sky}{HTML}{CEFFFF}
\definecolor{indigo}{HTML}{000066}
\definecolor{myblue}{HTML}{056FB1}
\definecolor{maroon}{HTML}{db0000}
\definecolor{forest}{HTML}{06961c}
\definecolor{lime}{HTML}{fcffad}
\definecolor{mygreen}{HTML}{024f16}
\definecolor{newgreen}{HTML}{aaf50a}
\definecolor{ocre}{HTML}{eb6315}

\usepackage{hyperref}
\usepackage{prettyref}
\newcommand{\pref}{\prettyref}
\newrefformat{eq}{equation (\ref{#1})}
\newrefformat{sec}{section~\ref{#1}}
\usepackage{bbold}

\newtcolorbox[auto counter,number within=section]{example}[2][]{ title style={right color=myblue!20, left color=myblue},
	colback=mylavender,colframe=newgreen!5,interior style={left color=newgreen!50,right color=white},fonttitle=\bfseries,breakable,enhanced jigsaw,
	title=Remark~\thetcbcounter: #2,#1}

\newtcolorbox{side}[1]{colback=lime,
	colframe=lime!20,interior style={left color=lime,right color=white},breakable,enhanced jigsaw}
\newtcolorbox{pozor}[1]{enhanced,colback=white,
	colframe=soft!20, interior style={left color=red!15,right color=white},fonttitle=\bfseries,breakable,enhanced jigsaw,
	title=#1}
\newtcolorbox[]{definition}[1][]{enhanced,borderline west={3pt}{-3pt}{cyan},colback=white,colframe =white, interior style={left color = ocre!5, right color =white}, breakable,enhanced jigsaw}

\renewcommand{\d}{\mathrm{d}}

\hypersetup{
	colorlinks=true,
	linkcolor=ocre,
	filecolor=cyan,      
	citecolor=ocre,
    urlcolor=myblue
}
\usepackage{tensor}
\usepackage{comment}
\usepackage{parskip} 
\usepackage{mathrsfs}
\usepackage{cleveref}

\begin{document}
\begin{samepage}
		\begin{flushleft}\huge{\textbf{Generalised Symmetries and Manifest Duality II: Curved Spacetime}}\end{flushleft}
		\vspace{20pt}
		{\color{myviolet}\hrule height 1mm}
		\vspace*{10pt}
		\begin{flushleft}
		\large 	\textbf{Subhroneel Chakrabarti}$\,{}^a$,  \textbf{Arkajyoti Manna}$\,{}^{b}$, \textbf{Madhusudhan Raman}$\,{}^c$
		\end{flushleft}

		\begin{flushleft}
			\emph{\large ${}^a$ Department of Theoretical Physics and Astrophysics, Faculty of Science, Masaryk University, 611 37 Brno, Czech Republic.}
			\\ \vspace{1mm}
            \emph{\large ${}^b$Center for Fundamental Physics, School of Physical Science and Technology,\\
ShanghaiTech University, 393 Middle Huaxia Road, Shanghai 201210, China}
			\\ \vspace{1mm}
            \large ${}^c$ \emph{Department of Physics, Ashoka University\\ Rajiv Gandhi Education City, Sonipat 131 029, India}
			\\ \vspace{3mm}
             \href{mailto:subhroneelc@sci.muni.cz}{subhroneelc@sci.muni.cz},
             \href{mailto:amanna@shanghaitech.edu.cn}{amanna@shanghaitech.edu.cn},
             \href{mailto:mraman@physics.du.ac.in}{madhusudhan.raman@gmail.com} \\
		\end{flushleft}

  \begin{flushright}
		\end{flushright}
		\section*{Abstract}
		{
			We extend the potential-based formulation of self-dual and duality-symmetric gauge theories developed in \cite{Chakrabarti:2025gyt} to curved spacetime. The gravitational coupling is encoded by the same metric-dependent linear map of Sen's formalism. The parent new action retains the higher-form $h$-gauge symmetry, and
different gauge choices reproduce either Sen's flux-based formulation or a
potential-based description in which the shadow sector decouples directly at the
level of the action. We also derive explicit form of the gravitational map for the toroidally
reduced theory in $D=4n$ and for a polyform formulation in arbitrary dimension,
both of which apply equally to ours as well as Sen's formalism. As non-trivial checks, we
reproduce the \'Alvarez-Gaum\'e-Witten gravitational anomalies of the two-dimensional chiral boson
without fermionisation, and of the ten-dimensional self-dual five-form. Finally,
on $\mathrm{AdS}_5\times S^5$ the physical on-shell action yields the
non-vanishing boundary contribution required by holography without adding a
boundary term by hand. We also establish a local on-shell identification with the
textbook supergravity RR four-form potential.
            }
\end{samepage}
	\newpage
	\vspace{10pt}
	\noindent\rule{\textwidth}{1pt}
	\pagecolor{white}
    \setcounter{tocdepth}{1}
	\tableofcontents\thispagestyle{fancy}
	\noindent\rule{\textwidth}{1pt}
	\vspace{10pt}

\section{Introduction} \label{sec:intro}

In an earlier paper \citep{Chakrabarti:2025gyt}, we introduced a new formulation of self-dual and duality-symmetric gauge theories in flat spacetime. The central idea was to formulate these theories in terms of both gauge potentials \textit{and} fluxes. The resulting framework contains both a gauge potential and a self-dual field related by a higher-form gauge symmetry we dubbed $h$-gauge symmetry. Different gauge choices were then shown to lead either to Sen's formulation or to a description in which the physical degrees of freedom are encoded entirely in gauge potentials. In this latter form, the shadow sector decouples directly at the level of the action, while the standard coupling of gauge potentials to conserved currents is preserved. This earlier paper also extended the construction to electric-magnetic duality in general dimensions and established the equivalence of the various gauge-fixed descriptions via explicit calculations.

The purpose of the present paper is to extend this framework to curved spacetime. This is a necessary step if the formalism is to be applied to gravitational backgrounds, quantum anomalies, and supergravity. The self-dual RR five-form of type IIB supergravity gives a concrete setting in which this extension is needed. Its self-duality condition depends on the physical metric, and its contribution to the gravitational anomaly provides a quantum test of the coupling \citep{Alvarez-Gaume:1983ihn}. The on-shell action of type IIB theory on $\mathrm{AdS}_5\times S^5$ provides a further test, since its value is constrained by holography. Some other physical questions where this extension will prove to be helpful are $M5$-brane worldvolume theory in curved spacetime, $6D$ supergravity that contains a self-dual flux in its tensor multiplet, and gravitational aspects of any EM-duality invariant theory, such as one in presence of both electrically and magnetically charged objects. These are questions about the gravitational response of the theory which cannot be settled by its flat-space equations alone.

A key observation is that the metric-dependent map in Sen's formulation \citep{Sen:2015nph,Sen:2019qit, Andriolo:2020ykk,Hull:2023dgp} can be incorporated into the new action without modification\footnote{For some recent applications of this map in various cases see \citep{Janaun:2024wya,Vanichchapongjaroen:2025psm,Lambert:2026ual}.}. Since this map depends only on the relation between the reference and physical metrics, it does not distinguish between a self-dual field treated as a fundamental flux and one constructed from gauge potentials. As a result, the curved-space extension follows naturally from the flat-space theory developed earlier.

The resulting action describes self-dual fields propagating on a general curved background while retaining the main advantages of the potential formulation. In particular, the shadow sector remains decoupled, and the physical equations of motion take the expected covariant form. The higher-form gauge symmetry continues to play an important role. One gauge choice reproduces Sen's curved-space action, while another leads to a formulation written entirely in terms of ordinary gauge potentials. In this gauge, external sources couple in the familiar way, and both fields and currents transform as standard differential forms leading to a manifestly diffeomorphism invariant action for the physical sector. The non-trivial coupling to gravity is encoded entirely through the metric-dependent map. We also formulate the theory using an arbitrary reference metric, adopting the recent two-metric construction of \citep{Hull:2023dgp}. This provides a more flexible framework in which the flat reference metric appears as a special case.

Having established the curved-space theory for self-dual case in $4n+2$-dimensions, we then investigate its implications for duality-symmetric systems in $4n$ and generic dimensions. Upon compactifying the self-dual theory on a two-torus, we produce a manifestly electric-magnetic duality-symmetric theory in $D=4n$. We derive the corresponding gravitational map governing the coupling to curved backgrounds and express it directly in terms of the reference and physical Hodge operators. We then generalise the construction further by combining a field strength and its dual into a self-dual polyform doublet. This leads to a unified curved-space description of duality-symmetric Abelian gauge theories in arbitrary spacetime dimension. To our knowledge, the explicit gravitational maps obtained in these reduced and general-dimensional settings have not appeared previously in the literature. These explicit maps are equally useful for our new formalism as well as Sen's formalism.

We then go on to apply our new formalism for two concrete physical tests.
The first test concerns gravitational anomalies. Since the coupling to gravity is encoded in a rather unconventional way through the metric-dependent map, it is important to verify that the quantum theory reproduces known anomaly structures \citep{Alvarez-Gaume:1983ihn}. We therefore compute gravitational anomalies directly within the potential-based formalism. For the two-dimensional chiral boson we recover the anomalous stress-tensor Ward identity \textit{without} resorting to a fermionisation, and for the ten-dimensional self-dual field we reproduce the celebrated \'Alvarez-Gaum\'e--Witten gravitational anomaly \citep{Alvarez-Gaume:1983ihn}. These results provide a stringent quantum check of the curved-space construction and confirm that it captures the correct gravitational response of self-dual fields.

Our second application concerns type IIB supergravity on $\mathrm{AdS}_5\times S^5$. This example is particularly interesting because it exposes a long-standing tension between self-dual formulations and holography. The holographic computation requires a non-vanishing ten-dimensional on-shell action proportional to the regularised volume of $\mathrm{AdS}_5$. However, both the conventional type IIB pseudo-action and Sen's flux-based action vanish when evaluated on shell. A modification proposed for the type IIB pseudo-action in \citep{Kurlyand:2022vzv} showed that one needs to add a pure boundary term. This motivated our previous work based on Sen's formulation \citep{Chakrabarti:2022jcb}, where too the correct result could only be obtained after adding an additional total-derivative contribution by hand. 

One of the most striking features of the potential-based formulation is that the required contribution emerges automatically. After removing the decoupled shadow sector, the physical action reduces on shell to a boundary term whose value is determined entirely by the physical fields. When evaluated on $\mathrm{AdS}_5\times S^5$, this boundary contribution reproduces precisely the non-zero result required by holography. Unlike previous approaches, no additional boundary prescription or supplementary term need be introduced. The correct answer is already encoded in the action itself. We regard this as strong evidence that the potential-based formulation is more synergistic with holography than existing flux-based descriptions.

The paper is organised as follows. In \cref{sec:Sen} we review the properties of the gravitational map and Sen's action in curved space needed throughout the paper.  We then construct in \cref{sec:new_SD} the curved-space form of the new action, including couplings to sources, its two-metric extension, and its relation to the textbook type IIB RR potential. In \cref{sec:D=4n} we study the dimensional reduction to $D=4n$ and derive the corresponding duality-symmetric gravitational map. In \cref{sec:general_D} we extend the construction to arbitrary spacetime dimension using polyform doublets. In \cref{sec:anomaly_calc} we compute gravitational anomalies and show that the formalism reproduces the expected quantum results in $D=2$ and $D=10$. In \cref{sec:ads_physical_onshell} we evaluate the physical on-shell action on $\mathrm{AdS}_5\times S^5$ and demonstrate its agreement with holographic expectations. We conclude in \cref{sec:conclusion}. We will predominantly use the form notation, however, we provide a dictionary to translate the equations into component notations in appendix~\pref{app:component_dictionary}. Some relevant details regarding construction of the gravitational map are given in appendix~\pref{app:construction_M} and \pref{app:4d_torus_complex_modulus}.


\section{Review of Sen's Action in a Curved Background} \label{sec:Sen}

Let us briefly recall Sen's action on curved spacetimes. For more details we refer the reader to \citep{Sen:2015nph,Sen:2019qit,Andriolo:2020ykk}. In this section we work on a $(4n+2)$-dimensional Lorentzian spacetime equipped with a dynamical metric which we will denote as $g$, and our goal will be to describe the dynamics of a self-dual $(2n+1)$-form field-strength $\mathcal{F}_{(g)} := \star_g \mathcal{F}_{(g)}$. 

\subsection{Without Sources}\label{sec:without_sources}
Sen's action for self-dual form fields in a gravitational background and in the absences of sources is
\begin{equation} \label{sen_action_without_source}
    S = \frac{1}{2}\int \d B \wedge \star \d B - \int \d B \wedge \mathcal{F} + \frac{1}{4}\int \mathcal{F} \wedge M(\mathcal{F}) \ .
\end{equation}
Here $\mathcal{F} = \star\mathcal{F}$ is a self-dual $(2n+1)$-form, but unusually it is self-dual with respect to the flat Minkowski metric $\eta$. The Hodge dual with respect to the Minkowski metric will be denoted $\star$. The physical curved metric is denoted as $g_{\mu\nu}$ and the Hodge star with respect to it will be denoted as $\star_g$.

A direct replacement of $\star$ by $\star_g$ in the kinetic terms would couple the shadow sector to gravity. Since $g$ is dynamical, this would also introduce interactions between the shadow and physical sectors, contrary to the required decoupling \citep{Sen:2015nph,Sen:2019qit}. Integrating out the shadow field in flat spacetime before coupling to gravity does not avoid this problem. As discussed in section~9 of the part I \citep{Chakrabarti:2025gyt}, the resulting Maxwell-like kinetic term for the $2n$-form potential $C$ cannot be coupled by the replacement
\begin{equation}
    -\frac{1}{2}\int \d C\wedge\star\d C
    \quad\longrightarrow\quad
    -\frac{1}{2}\int \d C\wedge\star_g\d C\,.
\end{equation}
The stress tensor of the expression on the right receives contributions from both $\d C+\star_g\d C$ and $\d C-\star_g\d C$. This prescription therefore couples the anti-self-dual part to gravity as well, and no longer describes the gravitational response of a self-dual field alone.

Sen's prescription instead keeps the reference Hodge star in the kinetic terms and introduces the physical metric through a separate linear map $M$. The map is algebraic in the metrics and contains no curvature tensors. We shall refer to this as a non-standard gravitational coupling to distinguish it from the textbook meaning of non-minimal coupling. It is chosen so that $\mathcal{F}-M(\mathcal{F})$ is self-dual with respect to $g$, while the shadow sector remains independent of $g$. The map has the following properties \citep{Andriolo:2020ykk,Hull:2023dgp}:

\begin{enumerate}\label{M-properties}
    \item $M$ is a linear map from $\eta$-self-dual to $\eta$-anti-self-dual forms, i.e.
    \begin{equation}
        \text{if} \quad \mathcal{F} = \star\mathcal{F}, \quad M(\mathcal{F}) = -\star M(\mathcal{F}).
    \end{equation}

    \item The quantity $\left[\mathcal{F} - M(\mathcal{F})\right]$ is by construction $g$-self-dual, i.e.
    \begin{equation}
        \mathcal{F} - M(\mathcal{F}) = \star_g\left[\mathcal{F} - M(\mathcal{F})\right]
    \end{equation}
    We will denote $\mathcal{F} - M(\mathcal{F}) := \mathcal{F}_{(g)}$.

    \item The map $M$ is symmetric in the following sense:
    \begin{equation}
        \mathcal{F}_1 \wedge M(\mathcal{F}_2) = \mathcal{F}_2 \wedge M(\mathcal{F}_1)
    \end{equation}

    \item Finally, we choose the action of $M$ on $\eta$-anti-self-dual forms to identically vanish.
\end{enumerate}

It was shown in \citep{Sen:2015nph} that such an $M$ can be explicitly constructed given a metric $g_{\mu\nu}$ and an equivalent, but alternate prescription was given in \citep{Andriolo:2020ykk} which is what we will also use. To keep the paper self-contained, we give a short summary of this construction in appendix~\prettyref{app:construction_M}. Indeed, \citep{Hull:2023dgp} extended the latter construction such that instead of $\eta$ we can use any reference metric $\hat{g}_{\mu\nu}$.

The equation of motion from (1) gives,
\begin{align}
    \delta_B S = 0 &\implies -\d{\star}\d B + \d\mathcal{F} = 0, \\
    \delta_\mathcal{F} S = 0 &\implies \frac{1}{2}(\d B - {\star}\d B) - \frac{1}{2}M(\mathcal{F}) = 0.
\end{align}

Defining $\hat{H} := \d B + \star \d B - \mathcal{F}$ and the definition of $\mathcal{F}_{(g)}$, we can re-write the EOM as
\begin{align}
    \d \hat{H} &= 0, \\
    \d \mathcal{F}_{(g)} &= \d \star_g \mathcal{F}_{(g)} = 0.
\end{align}

We recognise the latter as the derived free equation of motion for a $g$-self-dual flux. The former on the other hand is a decoupled $\eta$-self-dual field that is decoupled from all physical fields.

\subsection{With Sources}\label{sec:sen_action_with_source}

The extension to the case with sources is straightforward. However, since these sources are sources for fluxes, they possess one higher rank than the currents in textbook examples.

Let us denote by $\Omega$ a $(2n+1)$-form that acts as a classical source for $\mathcal{F}$. The action is given by
\begin{equation}
    S[\Omega] = \frac{1}{2}\int \d B \wedge {\star}\d B - \int \d B \wedge \mathcal{F} + \frac{1}{4}\int \left[\mathcal{F} + \Omega_+\right] \wedge M(\mathcal{F}+\Omega_+) + \int \mathcal{F} \wedge \Omega_- - \int \Omega_+ \wedge \Omega_-
\end{equation}
where $\Omega_\pm := \frac{1}{2}(1 \pm \star)\Omega$.

This leads to the following equations of motion for the physical sector:
\begin{equation}
    \d\mathcal{F}^\Omega_{(g)} = \d \star_g \mathcal{F}^\Omega_{(g)} = \d \Omega \ ,
\end{equation}
with $\mathcal{F}^\Omega_{(g)} := \mathcal{F} + \Omega_+ - M(\mathcal{F}+\Omega_+)$.

The $(2n+1)$-form $\Omega$ is related to the standard current $j$ in local patches, as
\begin{equation}
    \d^\dagger \Omega = j \qquad \left[\text{with } \d^\dagger := \star_g \, \d \,\star_g\right].
\end{equation}

The action is diffeomorphism invariant, but $B$ and $\mathcal{F}$ both possess non-standard transformation properties under diffeomorphism. A quick way to identify the correct diffeomorphisms is the following pair of observations:
\begin{enumerate}
    \item Extra field $\hat{H}$ has no ``knowledge'' of gravity. Therefore, it should be invariant under diffeomorphisms:
    \begin{equation}
        \delta_\xi \hat{H} = 0 \implies \delta_\xi(\d B + {\star}\d B) = \delta_\xi \mathcal{F}.
    \end{equation}

    \item The flux $\mathcal{F}$ should transform in such a way that up to boundary terms
    \begin{equation}
        \delta_\xi S[\Omega] = 0 \ .
    \end{equation}
\end{enumerate}
As a consequence, one can check that on-shell, the physical flux has the expected diffeomorphism properties, i.e.
\begin{equation}\label{diffeo_standard}
    \delta_\xi \mathcal{F}_{(g)} = -\pounds_\xi \mathcal{F}_{(g)}
\end{equation}
where $\pounds_{\xi}$ is the Lie derivative along $\xi^\mu(x)$.

\section{New Action in Curved Spacetime} \label{sec:new_SD}

The new action in the potential based formalism of \cite{Chakrabarti:2025gyt} for self-dual fields in $(4n+2)-$dimensional curved spacetime sans sources is given by
\begin{align}
    S[B,C,\Sigma,g] &= \frac{1}{2} \int \d B \wedge \star \d B - \int \d B \wedge F + \frac{1}{4} \int F \wedge M(F) \; , \\
    F &:= \d C + \star \d C - \Sigma \quad \mathrm{with} \quad \Sigma := \star \Sigma \;.
\end{align}
The $2n$-form field $B$ is the extra ``shadow'' field, just like in Sen's formalism. The self-dual field-strength is captured entirely by the variable $F$ which is built out of $2n$-form potential-like field $C$ and an $\eta$-self-dual flux $\Sigma$. The map $M$ is the same one described in the previous section. In particular, note that the construction of $M$ is only contingent upon the metric $g_{\mu \nu}$. 

It was shown in \citep{Chakrabarti:2025gyt} with a rigorous Hamiltonian analysis that as long as the interaction term is independent of the extra field $B$, which is clearly the case for the gravitational coupling here as well, the shadow sector decouples completely from the physical sector at the level of full quantum Hamiltonian.

Additionally, the new action has a higher-form gauge invariance, that was called $h$-gauge symmetry, under which 
\begin{equation}
    \delta_h C = \d_\eta^\dagger h \quad \mathrm{and} \quad \delta_\Sigma = \Delta_\eta h \quad \mathrm{with} \quad h =\star h \;,
\end{equation}
and all other fields are invariant. We have denoted by $\d_\eta^\dagger = \star \d \star$ and similarly $\Delta_\eta$ is the Hodge Laplacian with respect to $\eta$. This gauge symmetry leaves $F$ and the action invariant. It was shown in \citep{Chakrabarti:2025gyt} that this $h$-gauge symmetry can be fixed to set either 
\begin{itemize}
    \item $(\d + \star \d) C =0$, which leads us to Sen's action.
    \item Or set $\Sigma = 0$, which leads us to a novel potential-based action. 
\end{itemize} 

In this paper, our focus is on the second case with the gravitational coupling. The partially gauge fixed, potential-based action without source therefore looks like
\begin{equation} \label{newaction}
    S[B,C,g] = \frac{1}{2} \int \d B \wedge \star \d B - \int \d B \wedge (\d C + \star \d C) + \frac{1}{4} \int (\d C + \star \d C) \wedge M(\d C + \star d C) \;,
\end{equation}
where the self-dual field-strength is now completely fixed by the gauge fixed version of $F$ which is now $F_c = \d C + \star \d C$.

Just like the flat spacetime case, we can ignore the total derivative term \footnote{Assuming appropriate boundary conditions for now.} redefine our fields $B \to B + C$, complete the square, and achieve a complete decoupling of the shadow sector at the level of co-ordinate space action itself
\begin{equation}
    S[B,C,g] = \frac{1}{2} \int \d B \wedge \star \d B - \frac{1}{2} \int \d C \wedge \star \d C + \frac{1}{2} \int \d C \wedge M(\d C) \;.
\end{equation}
It is easy to check that the physical sector equation of motion is now 
\begin{equation}
    \d [F_c - M(F_c)] = \d F^{(g)}_c = \d \star_g F^{(g)}_c = 0 \;,
\end{equation}
as expected of a free $g$-self-dual field-strength $F^{(g)}_c$.

\subsection{Turning on Sources}

The advantage of the potential-based formalism over Sen's is most transparent when we turn on the sources. We also want to keep the situation of type IIB SUGRA in mind, where in the presence of sources, the $g$-self-dual field-strength gets twisted by the source term as well.

It was pointed out in \citep{Chakrabarti:2025gyt} that since the free theory has no propagating anti-self-dual part, we can actually now turn on a standard source via the familiar current $2n$-form that we will denote by $J$. This will source an anti-self-dual part as well, but since this anti-self-dual part has no kinetic term, it will not propagate and contribute to any physical observable. The action with the sources take the form
\begin{align}\label{newaction_with_source}
    S[B,C,J] = \frac{1}{2} \int \d B \wedge \star \d B - \frac{1}{2} \int \d (C+J) \wedge \star \d (C+J) \cr \quad + \frac{1}{2} \int \d (C+J) \wedge M(\d (C+J)) 
    - \int C \wedge \star_g J\,.
\end{align}
Defining the twisted $\eta$-self-dual variable
\begin{align}
    \mathcal{F}^{J} :=\d (C+J) + \star \d (C+J)\,,
\end{align}
 it is easy to check the equation of motion obtained by varying the action with respect to $C$ gives 
\begin{align}
    \d \left(\mathcal{F}^{J}-M(\mathcal{F}^{J}) \right) = \star_g J\,.
\end{align}
Using the properties of the map $M$ mentioned in \cref{sec:without_sources}, the combination $(\mathcal{F}^{J}-M(\mathcal{F}^{J}))$ is self-dual w.r.t $g$ and we recognize it to be the physical $g$-self-dual flux
\begin{align}
    \mathcal{F}^{J}_{(g)} := (\mathcal{F}^{J}-M(\mathcal{F}^{J})) \,, \quad \star_g  \mathcal{F}^{J}_{(g)} =  \mathcal{F}^{J}_{(g)}\,.
\end{align}
Thus we can re-express the equation of motion as
\begin{align}
    \d \mathcal{F}^{J}_{(g)} = \star_g J\,,
\end{align}
which is the correct equation of motion satisfied by a self-dual field in curved background with metric $g$. Here again the dynamics of $(2n)$-form shadow field $B$ completely decouples from the physical $(2n)$-form field $C$, as it ought to.

\subsection{Diffeomorphisms}
We shall consider both the physical field $C$ and the rank $2n$ background source field $J$ as true forms defined on the manifold with the curved metric $g$, i.e. they transform under the general coordinate transformation as in \cref{diffeo_standard}. The matter coupling between $C$ and $J$ in \cref{newaction_with_source}, is then manifestly diffeomorphism invariant and the \textit{minimal} coupling with gravity is now replaced by a \text{non-standard} coupling in the kinetic term of the action via the map $M$, defined in \cref{sec:without_sources}.

For the rest of this section, we focus on checking the diffeomorphism invariance of the rest of the action. This is non-trivial, since, even though $C$ and $J$ are true differential forms with respect to the curved metric, the $\eta$-self-dual field $\mathcal{F}^{J}$ is not.

To see that the action is invariant under the spacetime diffeomorphisms, let us express it in terms of $\mathcal{F}^J$ in the original form, i.e.~after restoring the total derivative term and undoing the field redefinition $B \to B + C$. We get,
\begin{align}
    S[B,C] \rightarrow \frac{1}{2} \int \d B \wedge \star \d B - \int \d B \wedge \mathcal{F}^J +\frac{1}{4} \int \mathcal{F}^J \wedge M(\mathcal{F}^J) \,.
\end{align}
This action is exactly same structure as in \cref{sen_action_without_source} and thus invariant under the same non-standard diffeomorphism discussed in \cref{sec:sen_action_with_source}. The only difference is that we no longer need to independently specify the diffeomorphism transformation of $\mathcal{F}^J$. It follows immediately from its definition in terms of $C$ and $J$ and the data that $C$ and $J$ transform like ordinary forms. A benefit of this, is that in this potential-based action, we have even off-shell, 
\begin{equation}
    \delta_\xi \mathcal{F}^J = - \pounds_\xi \mathcal{F}^J \;.
\end{equation}
So the the action is indeed diffeomorphism invariant, with much simpler diffeomorphism rules for the various physical field and the sources. It is also clear, that the non-standard diffeomorphism rules enter via two fronts:
\begin{enumerate}
    \item The extra field must not couple to gravity --- this specifies their transformation rules.
    \item The non-standard coupling to gravity implies that the physical flux must be written as $\mathcal{F}^J - M(\mathcal{F}^J)$, with $\mathcal{F}^J$ having non-standard transformation due to being an $\eta$-self-dual form. 
\end{enumerate} 

Indeed, for source-free case we can reach a manifestly diffeomorphism invariant form of the physical part of the action. The complete action in shifted variables is
\begin{equation}\label{eq:complete_shifted_action}
    S
    =
    S_{\mathrm{sh}}[B]
    +
    S_{\mathrm{phys}}[C]
    + S_{\mathrm{bdy}}[B,C] \;.
    \end{equation}
Focusing only on $S_{\mathrm{phys}}[C]$, which has the form
\begin{align}
    S_{\mathrm{phys}}[C]
    &=
    -\frac{1}{2}
    \int_{\mathcal M}
    \d C\wedge\star \d C
    +
    \frac{1}{2}
    \int_{\mathcal M}
    \d C\wedge M(F)
    \nonumber\\
    &=
    -\frac{1}{2}
    \int_{\mathcal M}
    \d C\wedge
    \left[
        \star \d C-M(F)
    \right].
\end{align}
Since
\begin{equation}
    F_{(g)}
    =
    F-M(F)
    =
    \d C+\star \d C-M(F),
\end{equation}
we have
\begin{equation}
    \star \d C-M(F)=F_{(g)}-\d C.
\end{equation}
Using $\d C\wedge \d C=0$, on substitution, we obtain the manifestly diffeomorphism invariant action
\begin{equation}\label{eq:physical_action_compact}
    S_{\mathrm{phys}}[C]
    =
    -\frac{1}{2}
    \int_{\mathcal M}
    \mathrm{d}C\wedge F_{(g)} \ .
\end{equation}

\subsection{Extension to Two-Metric Form} \label{sec:two_metric}

In \citep{Hull:2023dgp}, Hull extended Sen's formulation by promoting the reference
metric from $\eta$ to an arbitrary Lorentzian metric $\hat{g}$. The potential-based
action admits an identical extension. Let $\widehat{\star} := \star_{\hat{g}}$,
and note that $\widehat{\star}^{\,2} = +1$ on $(2n+1)$-forms in Lorentzian
signature, so the chiral projectors
$\widehat{P}_{\pm} := \frac{1}{2}(\mathbb{1} \pm \widehat{\star})$ are well
defined. The two-metric action is obtained by replacing every $\star$ by
$\widehat{\star}$, so we write
\begin{align}\label{two_metric_action}
    S[B,C,\Sigma;\hat{g},g] &= \frac{1}{2}\int \d B \wedge \widehat{\star}\, \d B
    - \int \d B \wedge F
    + \frac{1}{4}\int F \wedge M(F)\,, \cr
    F &:= \d C + \widehat{\star}\, \d C - \Sigma \quad \mathrm{with} \quad
    \Sigma = \widehat{\star}\, \Sigma\,,
\end{align}
where $M$ is now a linear map from $\hat{g}$-self-dual to
$\hat{g}$-anti-self-dual forms, satisfying the properties of
\cref{sec:without_sources} with $\eta \to \hat{g}$. In particular,
$F_{(g)} := F - M(F)$ is $g$-self-dual and $M$ annihilates
$\hat{g}$-anti-self-dual forms. The $h$-gauge symmetry persists,
\begin{equation}
    \delta_h C = \d^{\dagger}_{\hat{g}} h\,, \qquad
    \delta_h \Sigma = \nabla_{\hat{g}} h\,, \qquad
    h = \widehat{\star}\, h\,,
\end{equation}
with $\d^{\dagger}_{\hat{g}} := \widehat{\star}\,\d\,\widehat{\star}$ and
$\nabla_{\hat{g}} := \d\,\d^{\dagger}_{\hat{g}} + \d^{\dagger}_{\hat{g}}\,\d$,
since the computation establishing $\delta_h F = 0$ uses only
$\widehat{\star}^{\,2} = 1$ and $\d^{\dagger}_{\hat{g}} = \widehat{\star}\,\d\,\widehat{\star}$.
The two gauge fixings of \citep{Aggarwal:2025fiq} now interpolate between
Hull's action \citep{Hull:2023dgp}, obtained by setting
$(\d + \widehat{\star}\,\d)\,C = 0$, and its potential-based counterpart,
obtained by setting $\Sigma = 0$ with
$F_c = \d C + \widehat{\star}\,\d C$.

The construction of $M$ detailed in \citep{Hull:2023dgp} is purely algebraic and pointwise, requiring only the
pair $(\hat{g},g)$. It is also a natural generalisation of the construction of \citep{Andriolo:2020ykk} with which it coincides whenever $\hat{g} = \eta$. We briefly summarise the result here and refer to the original paper for details and proofs.

Let $\hat{\lambda}^{I}_{\pm}$ be bases of
$\hat{g}$-(anti-)self-dual $(2n+1)$-forms. Here $I$ runs over the number of basis for (anti)-self-dual $(2n+1)$-forms in $(4n+2)$ dimensions. The basis for $g$-self-dual forms $\Lambda^{I}_{+}$ can be expanded locally as
\begin{equation}\label{two_metric_SA}
    \Lambda^{I}_{+} = \widehat{\mathcal{S}}^{I}{}_{J}\, \hat{\lambda}^{J}_{+}
    + \widehat{\mathcal{A}}^{I}{}_{J}\, \hat{\lambda}^{J}_{-}\,,
\end{equation}
which defines the matrices $\widehat{\mathcal{S}}$ and $\widehat{\mathcal{A}}$.
Setting $M = -\widehat{\mathcal{S}}^{-1}\widehat{\mathcal{A}}$ on
$\hat{g}$-self-dual forms, and zero otherwise, reproduces all the required
properties \citep{Andriolo:2020ykk,Hull:2023dgp}: for any $\hat{g}$-self-dual
$F$, the combination $F - M(F)$ is a linear combination of the
$\Lambda^{I}_{+}$ and hence $g$-self-dual. The map exists wherever
$\widehat{\mathcal{S}}$ is invertible. At $g = \hat{g}$ one finds
$\widehat{\mathcal{S}} = 2\cdot\mathbb{1}$ and $\widehat{\mathcal{A}} = 0$, so
$M$ is defined for all $g$ continuously connected to $\hat{g}$ without crossing
the degeneration locus at $\det \widehat{\mathcal{S}} = 0$ \citep{Hull:2023dgp}.

The structure of the dynamics is unchanged. The combination
$\widehat{H} := \d B + \widehat{\star}\,\d B - F$ is a free
$\hat{g}$-self-dual field, while the physical sector obeys
\begin{equation}
    \d F_{(g)} = \d \star_g F_{(g)} = 0\,.
\end{equation}
Since the gravitational coupling remains independent of $B$, the decoupling of
the shadow sector holds as before, now with respect to the background
$\hat{g}$ \citep{Chakrabarti:2025gyt,Hull:2023dgp}. The reference metric
therefore enters only the decoupled sector: the physical content, a closed
$g$-self-dual flux, is identical for every admissible choice of $\hat{g}$.
Sources are incorporated exactly as in \cref{newaction_with_source}, with
$\mathcal{F}^{J} = \d(C+J) + \widehat{\star}\,\d(C+J)$ and the matter coupling
$\int C \wedge \star_g J$ unmodified.

The two-metric form also inherits the simpler diffeomorphism structure just like Hull's extension of Sen's action. But more crucially, it allows us to not be reliant on a formalism that works with a specific reference metric, viz. $\eta$. The freedom to adapt the reference metric is very useful, especially in performing dimensional reductions. In the rest of this paper we will focus primarily on the case $\hat{g} = \eta$ unless otherwise specified.

\subsection{On-Shell Comparison with the Standard RR Potential} \label{sec:onshell_potential_map} 

Given the reappearance of potentials which are $2n$-forms on the spacetime manifold, it is natural to ask if these are the same as the gauge fields we encounter in textbooks. For specificity, we focus on $\hat{g}=\eta$ and specialise to the case of type IIB SUGRA in $D=10$ with the physical
flux of interest being the self-dual RR 5-form \footnote{All statements made here trivially extend to arbitrary
$(2n+1)$-forms in $D=4n+2$.}. Our goal is to relate the
potential $C$ of the new formalism to the standard RR potential $C^{(4)}$. Since the
conventional description of $C^{(4)}$ proceeds through a pseudo-action supplemented
by a self-duality constraint, rather than a genuine action, the two fields can
\textit{only} be compared on-shell, by identifying the physical $g$-self-dual flux evaluated
on classical solutions. 

We work in the absence of sources (the generalisation to the case with sources is straightforward) in this section. Define the projectors 
\begin{equation} 
P_\pm := \frac{1}{2}\left(\mathbb{1}\pm\star\right), \qquad P^g_\pm := \frac{1}{2}\left(\mathbb{1}\pm\star_g\right).
\end{equation} 
Following \citep{Andriolo:2020ykk}, it is useful to define the linear map
\begin{equation} 
m:=\mathbb{1}-M. 
\end{equation}
The map $m$ sends $\eta$-self-dual five-forms to $g$-self-dual five-forms. Moreover, every $g$-self-dual five-form $W$ satisfies the identity 
\begin{equation}\label{eq:m_reconstruction_RR} 
W=m\!\left(P_+W\right) \quad \text{if} \quad W = \star_g W.
\end{equation} 
The physical five-form flux in the standard textbook description is 
\begin{equation} 
F^{(5)}_{\mathrm{RR}} := 2P^g_+\d C^{(4)}, 
\end{equation} 
whereas in the potential-based formulation it is 
\begin{equation} 
F^{(5)}_{\mathrm{new}} := m\!\left(2P_+\d C\right). 
\end{equation} 
The on-shell matching condition is therefore \begin{equation}\label{eq:RR_flux_matching_simple} 
2P^g_+\d C^{(4)} \cong m\!\left(2P_+\d C\right). 
\end{equation} 

In the source-free type IIB theory, the standard RR field strength obeys the self-duality constraint
\begin{equation}\label{eq:RR_standard_selfduality} 
\d C^{(4)} \cong \star_g\d C^{(4)}. 
\end{equation} 
Consequently, 
\begin{equation} 
P^g_+\d C^{(4)} \cong \d C^{(4)}, 
\end{equation} 
and \eqref{eq:RR_flux_matching_simple} reduces to \begin{equation}\label{eq:RR_matching_reduced}
\d C^{(4)} \cong m\!\left(P_+\d C\right).
\end{equation}
Since $\d C^{(4)}$ is $g$-self-dual, the identity \eqref{eq:m_reconstruction_RR} also gives 
\begin{equation}\label{eq:RR_reconstruction_C4} 
\d C^{(4)} \cong m\!\left(P_+\d C^{(4)}\right). 
\end{equation} 
Subtracting \eqref{eq:RR_reconstruction_C4} from \eqref{eq:RR_matching_reduced}, we obtain 
\begin{equation} 
m\!\left[ P_+\left(\d C-\d C^{(4)}\right) \right] \cong 0. 
\end{equation} 
The restriction of $m$ to the $\eta$-self-dual subspace is injective. Indeed, for any $\eta$-self-dual five-form $\omega_+$, 
\begin{equation} 
P_+m(\omega_+) = P_+\left(\omega_+-M(\omega_+)\right) = \omega_+,
\end{equation} 
because $M(\omega_+)$ is $\eta$-anti-self-dual. Hence 
\begin{equation} 
m(\omega_+)=0 \qquad\Longrightarrow\qquad \omega_+=0.
\end{equation} 
It follows that 
\begin{equation}\label{eq:RR_difference_ASD} 
P_+\left(\d C-\d C^{(4)}\right) \cong 0. 
\end{equation} 
Therefore the difference between the two field strengths is $\eta$-anti-self-dual: 
\begin{equation}\label{eq:RR_rho_definition} 
\d C-\d C^{(4)} \cong \rho_-, \qquad \star\rho_-=-\rho_-. 
\end{equation} 

Since both $\d C$ and $\d C^{(4)}$ are exact, we must have
\begin{equation} \d\rho_-=0,
\end{equation}
and locally one may write
\begin{equation}\label{eq:RR_rho_exact}
\rho_-=\d\mu, \qquad P_+\d\mu=0.
\end{equation} 

The gauge-fixed potential formulation has the residual harmonic symmetry that can be also written as
\begin{equation}\label{eq:RR_residual_shift} 
C\longmapsto C+\nu, \qquad P_+\d\nu=0.
\end{equation} 
Choosing $\nu=-\mu$ removes the anti-self-dual remainder:
\begin{equation} 
\d C \longmapsto \d C-\d\mu \cong \d C^{(4)}.
\end{equation} 

Thus, locally and on-shell, one may choose the representative \begin{equation}\label{eq:RR_fieldstrength_dictionary}  
\d C\cong\d C^{(4)} .
\end{equation} 

It then follows that 
\begin{equation} \d\left(C-C^{(4)}\right)\cong0. 
\end{equation} 

Locally,
\begin{equation}
C-C^{(4)} \cong \d\Lambda^{(3)}, 
\end{equation} so an ordinary $U(1)$ gauge transformation may be used to set \begin{equation}\label{eq:RR_potential_dictionary}  
C\cong C^{(4)} . 
\end{equation}

The identification is local. Globally, closed but non-exact forms and non-trivial flux sectors may obfuscate the equality of the two potentials. We leave that discussion to future work.

\section{The New Action: $D=4n$ Case}\label{sec:D=4n}

In this section, for simplicity, we use only square metric in the bulk. The construction can be carried out using torus metric with more general complex structure. We present the details in appendix \cref{app:4d_torus_complex_modulus}.

The dimensional reduction of the first two terms in the new action in \cref{newaction} over a torus was carried out in our earlier work \cite{Chakrabarti:2025gyt}. Therefore, in this section we focus solely on the term in \cref{newaction} responsible for the gravitational coupling of the self-dual field in $(4n+2)$-dimensions
\begin{align}
    S_{4n+2} = \frac{1}{4} \int \mathcal{F} \wedge M(\mathcal{F})\,.
\end{align}
Focussing on the $2n$-forms after the torus reduction, the gravitational coupling in $4n$ dimensions can be expressed as
\begin{align}
    S_{4n} = \frac{1}{4} \int \mathbf{F} \wedge M (\mathbf{F})\,,\qquad \mathbf{F} = \begin{pmatrix}
        F\\
        -\widetilde{F}
    \end{pmatrix}\,,
\end{align}
where the rank-$2n$ field strength and its dual are given by
\begin{align}
    F &= \d A + \star \d \check{A} \,,\cr
    \widetilde{F} &= \star \d A - \d \check{A} \,,
\end{align}
using $\star^2 =-1$, with $\star$ now defined w.r.t the $4n$-dimensional flat metric. In the rest of this section, our goal would be to construct the map $M$ explicitly following \cite{Andriolo:2020ykk}. 

We use the metric of square torus for the compactification from $4n+2$ dimensions to $4n$ dimensions. Consider the torus metric to be\footnote{In general the square torus has the following metric $G_{ab}= A \delta_{ab}$ with $A= (2\pi R)^2$ where the compact directions are defined as $x^a \sim x^a + 2\pi R$.}
\begin{align}\label{square_torus}
    G_{ab} = \begin{pmatrix}
        1 & 0 \\
        0 & 1
    \end{pmatrix}\,,
\end{align}
and the $(4n+2)$-d metric has the form: $g_{4n+2} = g_{4n} \otimes G_2$.
Let us call the $T^2$ directions as $\{ y^a\}$ where the index $a$ running over $(4n+1)$ and $(4n+2)$. The Hodge star action along these directions is given by
\begin{align}
    \star_G dy^a = \Omega^a{}_b dy^b \,,
\end{align}
where 
\begin{align}\label{omega_matrix_reps_4n}
 \Omega^a{}_b =\begin{pmatrix}
     0 & 1\\
     -1 & 0
 \end{pmatrix} \,.
\end{align}
Thus the Hodge star with respect to the torus metric acts as 
\begin{align}\label{Hodge_star_torus}
    \star_{G_2} dy^1 = dy^2 \,, \qquad \star_{G_2} dy^2 = - dy^1 \,.
\end{align}
Let us introduce self-dual and anti-self-dual bases w.r.t the flat metric: $\lambda_{\pm}^{Ia}$ for a generic $(2n+1)$-form with $I$ denotes the indices of $2n$-form: $(\mu_1 \cdots \mu_{2n}) \forall$ $\mu_i \in[0,\cdots,4n-1]$.. However, since our goal is to construct the map $M$ after torus reduction, we adopt a particular choice for these bases \footnote{For the case of circle reduction, similar choice was adopted in \cite{Chakrabarti:2023czz}.}
\begin{align}
    \lambda_+^{Ia} & = e^I \wedge dy^1 + \tilde{e}^I \wedge dy^2\,,\cr
    \lambda_-^{Ia} & = e^I \wedge dy^1 - \tilde{e}^I \wedge dy^2\,,
\end{align}
where $e^I$ and $\tilde{e}^I = \star e^I$ are the bases of the $2n$-forms in $4n$ dimensions we get after the dimensional reduction. With these bases, the $2n$-forms can be expressed as 
\begin{align}
    F= F_I e^I \,, \qquad \widetilde{F}=\star F =\widetilde{F}_I \tilde{e}^I \,.
\end{align}

Next, we consider the basis $\Lambda_+^I$ which is self-dual w.r.t the metric $g_{4n+2}$
\begin{align}
    \Lambda_+^{Ia} = \star_{g_{4n+2}} \Lambda_+^{Ia} = (\star_{g_{4n}} \otimes \star_{G_2}) \Lambda_+^{Ia} \,.
\end{align}
Since this is a $(2n+1)$-rank form in $(4n+2)$ dimensions, we can express $\Lambda_+^{Ia}$ in terms of the bases $\lambda_{\pm}^{Ia}$ locally on a coordinate patch
\begin{align} \label{4n_bases_g}
    \Lambda_+^{Ia} & = \mathcal{S}^I{}_J \lambda_+^{Ja} + \mathcal{A}^I{}_J \lambda_-^{Ja}\cr
    & = (\mathcal{S} + \mathcal{A})^I{}_J \, e^J \wedge dy^1 + (\mathcal{S} - \mathcal{A})^I{}_J \, \tilde{e}^J \wedge dy^2 \,,
\end{align}
for some appropriate matrix coefficients $\mathcal{S}$ and $\mathcal{A}$. Given this map between the two basis, we can construct linear map as $M=-\mathcal{S}^{-1}\mathcal{A}$, satisfying all the properties in \cref{M-properties}. 
Focusing on the $2n$-form sector with $F$ and $\widetilde{F}$ after the reduction, a natural choice to express $\Lambda_+^{I}$ as follows \footnote{Equivalently, we can also choose to work with the $\widetilde{F}$ sector after the reduction. In this case, we can express it as $\Lambda_+^{Ia} = \tilde{e}^I \wedge dy^2 + \star_{g_{4n+2}}(\tilde{e}^I \wedge dy^2)$.
}
\begin{align}\label{int_calc_2}
     \Lambda_+^{Ia} = e^I \wedge dy^1 + \star_{g_{4n+2}}(e^I \wedge dy^1)\,.
\end{align}
Using the representation of $\Omega$ matrix in \eqref{omega_matrix_reps_4n}, the Hodge star in the second term can be written as
\begin{align}
    \star_{g_{4n+2}}(e^I \wedge dy^1)= (\star_{g_{4n}} e^I \wedge \star_{G_2} dy^1) = \star_{g_{4n}} e^I \wedge  dy^2\,.
\end{align}
Thus we get
\begin{align}
    \Lambda_+^{Ia} = e^I \wedge dy^1 + \star_{g_{4n}} e^I \wedge dy^2\,.
\end{align}
Comparing the above expression with \eqref{4n_bases_g}, we find
\begin{align}
    \mathcal{S} &= \frac{1}{2} \left( \mathbb{1}- \star_{g_{4n}} \star \right)\,,\cr 
    \mathcal{A} &= \frac{1}{2} \left(\mathbb{1}+ \star_{g_{4n}} \star \right)\,.
\end{align}
Thus the linear map $(M)$ of the dimensionally reduced theory is given by
\begin{align}
    M = - \left( \mathbb{1} - \star_{g_{4n}} \star \right)^{-1} \left( \mathbb{1} + \star_{g_{4n}} \star \right)\,.
\end{align}

\section{The New Action: General Case}\label{sec:general_D}

In general $D$ dimensions, other than $D=(4n+2)$, the distinction of self and anti-self dual fields ceases to exists. 
Thus to continue our analysis, we define a polyform $\mathbf{F}$ of rank $(p+1,D-p-1)$ with the following doublet structure 
\begin{align}\label{polyform_F}
    \mathbf{F} = \begin{pmatrix}
        F\\
        \widetilde{F}
    \end{pmatrix}\,,
\end{align}
where the $(p+1)$-form field strength $F$ and its Hodge dual $\widetilde{F}$ of rank $(D-p-1)$ are given in terms of potentials as follows
\begin{align}
    F = \d A + \star \d \check{A} \,, \quad \widetilde{F} = \star \d A + s\, \d \check{A}\,,
\end{align}
and the above expressions allow us to read off the ranks of the potentials $A$ and $\check{A}$. The Hodge $\star$ is defined w.r.t the $D$-dimensional flat metric $\eta_D$.
We denote $\star^2$ acting on a $(p+1)$-form in $D$-dimensions as $s = -(-1)^{(p+1)(D-p-1)}$, which can be either $\pm 1$. In terms of the polyforms, the new action in $D$-dimensions is given by
\begin{align}\label{new_action_D}
    S_D = \frac{1}{2} \int \mathbf{G}^T \wedge \star \mathbf{G} - \int \mathbf{G}^T \wedge \mathbf{F} + \frac{1}{4} \int \mathbf{F}^T \wedge M (\mathbf{F})\,,
\end{align}
where the extra fields $(B,\check{B})$ are of the same rank as the gauge potentials $(A,\check{A})$, packaged into a polyform $\mathbf{G}$ defined as
\begin{align}
    \mathbf{G} = \begin{pmatrix}
    \d B\\
    \d \check{B}
\end{pmatrix}\,.
\end{align}

The polyform defined in \cref{polyform_F}
 is self-dual w.r.t the following operator 
\begin{align}
    \widehat{\star} = \begin{pmatrix}
        0 & s \, \star \\
        \star & 0
    \end{pmatrix} =: \Omega_s \star \,.
\end{align}
Since $s^2=1$, it is clear that $\widehat{\star}\, \mathbf{F} = \mathbf{F}$ in $D$-dimensions. With the polyform $\mathbf{F}$, we have introduced the gravitational coupling in $S_D$ in \cref{new_action_D} as $\mathbf{F}^T \wedge M(\mathbf{F})$, 
where the map $M$ has the following properties:
\begin{itemize}
    \item $M: \Lambda^{(p)} \rightarrow \Lambda^{(D-p)}$, where $\Lambda^{(p)}$ is the set of $p$-forms in $D$ dimensions. So $M(F)$ is a $(D-p-1)$ form and $M(\widetilde{F})$ is a $(p+1)$ form.
    \item  $M$ is a linear map from $\widehat{\star}$-self-dual forms to $\widehat{\star}$-anti-self-dual forms, i.e.
    \begin{equation}
        \text{if} \quad \mathbf{F} = \widehat{\star}\,\mathbf{F}, \quad M(\mathbf{F}) = -\widehat{\star} M(\mathbf{F}).
    \end{equation}
    This condition in terms of $F$ and $\widetilde{F}$ translates to the following conditions: $\star M (F)=-M(\widetilde{F})$ and $s\star M(\widetilde{F}) = -M(F)$.
    \item The quantity $\left[\mathbf{F} - M(\mathbf{F})\right]$ is by construction $\widehat{\star}_g$-self-dual, i.e.\footnote{This is defined analogously as $ \widehat{\star}_{g_D}  := \Omega_s \star_{g_D}$.}
    \begin{equation}
        \mathbf{F} - \Omega_s M(\mathbf{F}) = \widehat{\star}_g\left[\mathbf{F} - \Omega_s M(\mathbf{F})\right]\,.
    \end{equation}
    We denote $\mathbf{F}_{(g)} := \mathbf{F} - \Omega_s M(\mathbf{F})$. 

    \item The map $M$ is symmetric in the following sense:
    \begin{equation}
        \mathbf{F}_1 \wedge M(\mathbf{F}_2) = \mathbf{F}_2 \wedge M(\mathbf{F}_1)\,.
    \end{equation}

   \item We choose the action of $M$ on $\widehat{\star}$-anti self-dual forms to identically vanish.
\end{itemize}

Next, we give an explicit construction of the linear map $M$ in general $D$-dimensions. Analogous to the $D=4n+2$ case, we introduce the $\widehat{\star}$ self-dual and anti self-dual bases
\begin{align}\label{hybride_bases_flat}
    \widehat{\lambda}^{I}_{\pm} = \begin{pmatrix}
        \lambda^{I } \\
        \pm \star \lambda^{I}
    \end{pmatrix} \Rightarrow \widehat{\star} \, \widehat{\lambda}^{I }_{\pm} = \pm \widehat{\lambda}^{I}_{\pm} \,.
\end{align}
The elements $\lambda^{I}$ and $\widetilde{\lambda}^{I}=\star \lambda^{I}$ serve as the bases for any generic rank $(p+1)$ and $(D-p-1)$-form respectively in $D$-dimensional flat spacetime.   
Here we use the same label $I$ to denote the indices of both $(p+1)$ and $(D-p-1)$-forms.
Let us write the gravitational analogue of $\widehat{\lambda}^{I}_+$ which is self-dual w.r.t $\widehat{\star}_{g_D}$ as follows
\begin{align}\label{hybride_bases_curved}
    \widehat{\Lambda}^{I}_+ = \begin{pmatrix}
        \lambda^{I } \\
        \star_{g_D}\lambda^{I }
    \end{pmatrix} \Rightarrow \widehat{\star}_{g_D} \, \widehat{\Lambda}^{I }_{+} = \widehat{\Lambda}^{I}_{+} \,.
\end{align}
By expressing $\widehat{\Lambda}^{I}_+$ in terms of the flat space basis locally on a patch as
\begin{align}
    \widehat{\Lambda}^{I}_+ = \widehat{\mathcal{S}}^I{}_J \widehat{\lambda}^{J}_+ + \widehat{\mathcal{A}}^I{}_J \widehat{\lambda}^{J}_- \,, \qquad \widehat{\mathcal{S}}= \begin{pmatrix}
        \mathcal{S} & 0 \\
        0 & \mathcal{S}
    \end{pmatrix}\,, \quad \widehat{\mathcal{A}}= \begin{pmatrix}
        \mathcal{A} & 0 \\
        0 & \mathcal{A}
    \end{pmatrix}
\end{align}
for some appropriate matrices $\mathcal{S}$,  $\mathcal{A}$, the linear map $M$ can be constructed as $M = -\widehat{\mathcal{S}}^{-1}\widehat{\mathcal{A}}$. 

Comparing \cref{hybride_bases_flat} and \cref{hybride_bases_curved} as a doublet equation 
\begin{align}
    \begin{pmatrix}
        \lambda^{I } \\
        \star_{g_D}\lambda^{I }
    \end{pmatrix}  = \begin{pmatrix}
        (\mathcal{S}^I{}_J + \mathcal{A}^I{}_J) \lambda^{J}\\  
         (\mathcal{S}^I{}_J - \mathcal{A}^I{}_J)\star \lambda^J
    \end{pmatrix}\,,
\end{align}
we find
\begin{align}
    \mathcal{S} + \mathcal{A} &= \mathbb{1}\,,\cr
    \mathcal{S} - \mathcal{A} &= s \star_{g_D} \star \,,
\end{align}
The matrices $ \widehat{\mathcal{S}}$ and $ \widehat{\mathcal{A}}$ can be determined using $ \mathcal{S}$ and $ \mathcal{A}$ and we obtain
\begin{align}
    \widehat{\mathcal{S}} &= \begin{pmatrix}
         \frac{1}{2} \left( \mathbb{1} + s \star_{g_D} \star \right) & 0\\
         0 &  \frac{1}{2} \left( \mathbb{1} + s \star_{g_{D}} \star \right)
    \end{pmatrix}\,,\cr 
    \widehat{\mathcal{A}} &= \begin{pmatrix}
         \frac{1}{2} \left(\mathbb{1} - s \star_{g_{D}} \star \right) & 0 \\
         0 &  \frac{1}{2} \left(\mathbb{1} - s \star_{g_{D}} \star \right)
    \end{pmatrix}\,.
\end{align}

\section{Gravitational Anomalies}\label{sec:anomaly_calc}

The anomalous stress-tensor Ward identities of \cite{Alvarez-Gaume:1983ihn} provide a concrete one-loop test that any proposed action for self-dual bosons in curved spacetime must satisfy. Reproducing the classical self-duality equation of motion from the action is not sufficient: the propagator and the gravitational interaction vertices must also give the anomaly of the physical chiral field. In the present formulation these vertices are determined by the map $M$, while the shadow sector does not couple to the physical metric. The examples below successfully test our prescription through the stress-tensor two-point function in two dimensions and the self-dual field contribution to the six-graviton amplitude in ten dimensions.

\subsection{Chiral Bosons in Two Dimensions}

\begin{figure}
	\centering
	\includegraphics{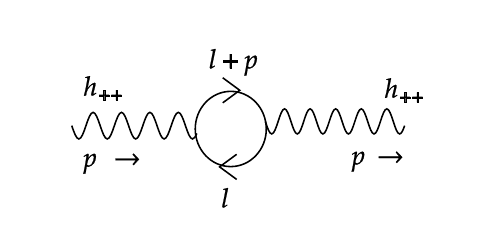}
	\caption{One-loop diagram contributing to gravitational anomaly of chiral boson.}
	\label{fig:chiral_1_loop}
\end{figure}

We start with the gravitational coupling
\begin{align}
    \mathcal{L}_{\text{int}} = \frac{\kappa}{2} h^{ab}T_{ab}\,,
\end{align}
where the flat space stress tensor of chiral boson is 
\begin{align}
    T_{ab}= \frac{1}{2} A_a A_b \,.
\end{align}
$A^a$ is the self-dual (chiral) field in two dimensions with $\{a,b,\cdots\}= 0,1$.
The self-duality condition in light-cone coordinates $x^{\pm}:= \frac{1}{\sqrt{2}}(x^0 \pm x^1)$ gives
\begin{align}
    A^a = \epsilon^{ab}A_b \Rightarrow A_- = 0 \,,
\end{align}
for $\epsilon^{01}=+1$. Thus the only non-zero component of the stress tensor is 
\begin{align}
    T_{++} = \frac{1}{2} A_+ A_+\,.
\end{align}
In our new formalism, by expressing 
\begin{align}
    A^a = (\partial^a + \epsilon^{ab} \partial_b) \varphi\,,
\end{align}
we can write the stress tensor in momentum space as
\begin{align}
    T_{++}(k) = 2 k^2_+ \varphi^2\,. 
\end{align}
Thus the three-point interaction vertex can be deduced from $\mathcal{L}_{\text{int}}$ as
\begin{align}
    V^{\varphi,\varphi}_{h_{++}}(k) = 2 \kappa k_+^2  \,.
\end{align}
Now let us consider the two-point function \cite{Alvarez-Gaume:1983ihn}
\begin{align}
    U(p) & = \int d^2x e^{ip\cdot x} \langle 0 | \mathcal{T} \left( T_{++}(x) T_{++}(0) \right)|0\rangle \,,
\end{align}
which is related to the one-loop amplitude in \cref{fig:chiral_1_loop}.
This amplitude can be computed straightforwardly 
\begin{align}
    U(p) &=  \int d\ell_+ d\ell_- V_{++}(l) V_{++}(\ell+p)  \frac{1}{\ell^2+i\epsilon} \frac{1}{(p+\ell)^2 +i\epsilon}\,,\cr
    & = 4 \kappa^2  \int d\ell_+ d\ell_- \frac{\ell_+}{\left[\ell_- + \frac{i\epsilon}{\ell_+}\right]}  \frac{(\ell+p)_+}{\left[\ell_- + p_- + \frac{i\epsilon}{(\ell+p)_+}\right]}
\end{align}
where we have used the standard scalar propagator $\langle \varphi (k)\varphi (-k)\rangle = \frac{1}{k^2+i\epsilon}$. Note that, as $\epsilon \rightarrow 0$, the two poles gives equal and opposite residues $\pm\frac{1}{p_-}$ and thus the integral vanishes.
The integral in $\ell_-$ is only non-zero when the two poles: $-i\epsilon/\ell_+$ and $-p_- - i\epsilon/(p+\ell)_+$ are on the opposite sides of real axis. Let us assume $p_+ >0$, then we shall restrict the $\ell_+$ integral to $0> \ell_+ > -p_+$. We get
\begin{align}
    U(p) = -4 \kappa^2 \int_{-p_+}^0 d\ell_+ \frac{(\ell_+ + p_+) \ell_+}{p_-} = \frac{2}{3}\kappa^2 \frac{p_+^3}{p_-}\,.
\end{align}
This implies $p_- U(p) \neq 0$ and violates the conservation of stress tensor at one-loop: $\partial_- T_{++}\neq 0$. This is the anomaly computed in \cite{Alvarez-Gaume:1983ihn} using fermionization of chiral boson in 2d.

\subsection{Gravitational Anomaly in Ten Dimensions}

In this section, we derive the self-dual propagator and graviton-self-dual field interaction vertex and show that they agree with the Feynman rules proposed in \cite{Alvarez-Gaume:1983ihn}. We use these to build the one-loop six-graviton amplitude\footnote{ For general $4n+2$ dimensions, we consider the one-loop diagram with $2n+2$ external gravitons.} and show that this loop integral is free of both ultraviolet and infrared divergences. This provides a non-trivial consistency check of the formalism- reproducing correct anomaly structure.

Let us begin with the two-point function of $C$ field. After gauge fixing using Lorentz covariant gauge fixing terms
\begin{align}
    S_{GF} = \frac{1}{2} \int \star \d \star B \wedge \d \star B - \frac{1}{2} \int \star \d \star C \wedge \d \star C\,,
\end{align}
we can obtain the two-point function involving $C$ field from the action in \cref{newaction} straightforwardly. In ten dimensions, the self-dual field is a rank-5 field strength and the potential $C$ in our new formalism is a rank-4 field with the following two-point function 
\begin{align}\label{two-point_CC}
    \langle C^{a_1\cdots a_4}(k) C^{b_1\cdots b_4}(k') \rangle = -\frac{i}{2 \times 4!} \delta^{(10)}(k+k') \frac{1}{k^2+i\epsilon} \eta^{a_1[b_1}\cdots \eta^{b_4]a_4} \,,
\end{align}
with $\{a,b,\cdots, a_i,b_i,\cdots\}= 0,\cdots, 9$ in this subsection.
Again, this two-point function is consistent with the field strength based perturbation theory. To see this, consider 
\begin{align}
  S & = S[B,F]+S_{GF}[B] \cr
   & = - \frac{1}{2} \int  B \wedge \star( \d^\dagger \d + \d \d^\dagger ) B - \int \d B \wedge F + \frac{1}{4} \int F \wedge M(F)\,.
\end{align}
Expanding the curved metric as $g_{ab} = \eta_{ab} + h_{ab}$, the term $F\wedge M(F)$ contributes at $\mathcal{O}(h)$. Therefore up to quadratic in $B$, we can write the action in momentum space
\begin{align}\label{action_F_mom}
    S =\frac{1}{4!} \int d^{10} k \left[ \frac{1}{2} B_{a_1\cdots a_4}(k) k^2 B^{a_1\cdots a_4}(-k) -i B^{a_1\cdots a_4}(-k) k^{c} F_{c a_1\cdots a_4}(k) + \mathcal{O}(h) \right]\,.
\end{align}
With the following field redefinition
\begin{align}
    B_{a_1\cdots a_4}(k) \to B'_{a_1\cdots a_4}(k) = B_{a_1\cdots a_4}(k) + \frac{i k^{c}}{k^2} F_{c a_1\cdots a_4}(k)\,,
\end{align}
the action in \cref{action_F_mom} can be brought into the form
\begin{align}
      S = \frac{1}{4!} \int d^{10} k \left[ B'_{a_1\cdots a_4}(k) k^2 B'^{a_1\cdots a_4}(-k) + F^{d a_1\cdots a_4}(-k)  \frac{k_d k^c}{k^2} F_{ca_1\cdots a_4}(k) + \mathcal{O}(h) \right]\,.
\end{align}
Thus the two-point function involving self-dual field $F_{A}(k)$ can be written as
\begin{align}\label{two-point_FF}
    \langle F^{A}(k) F^{B}(k') \rangle = - \frac{i}{5!}\delta^{(10)}(k+k') \frac{1}{k^2+i\epsilon} \mathcal{P}_+^{AC}K_{CD}(k) \mathcal{P}_+^{DB}\,,
\end{align}
where $(A,B, \cdots)$ indices denote collectively $(a_1\cdots a_5),(b_1 \cdots b_5),\cdots$ and so on. Here
\begin{align}
    K_{CD} \equiv K_{c_1\cdots c_5 d_1\cdots d_5} = k_{[c_1} k_{d_1} \eta_{c_2 d_2} \eta_{c_3 d_3} \eta_{c_4 d_4} \eta_{c_5 d_5]} \,,
\end{align}
which is totally anti-symmetric in both $\{c_i \}$ and $\{d_i\}$. The projection operators here are defined as
\begin{align}\label{projection_10d}
   \mathcal{P}_{\pm A}{}^B F_{B} & \equiv \mathcal{P}_{\pm\, a_1\cdots a_5}{}^{b_1\cdots b_5} F_{b_1\cdots b_5} \cr 
   & = \frac{1}{2 \times 5!} \left( \delta_{[a_1}{}^{[b_1} \cdots \delta_{a_5]}{}^{b_5]} \pm  \epsilon_{a_1\cdots a_5}{}^{b_1\cdots b_5} \right) F_{b_1\cdots b_5}\,.
\end{align}
Using the definition of the field strength in terms of the potential $C$ in momentum space 
\begin{align}\label{F_self_mom_6d}
    F_{a_1\cdots a_5} (k) & = \frac{i}{5!} \left( k_{[a_1} C_{a_2\cdots a_5]}(k) + \epsilon_{a_1\cdots a_5 b_1\cdots b_5} k^{b_1} C^{b_2\cdots b_5}(k) \right)\,,\cr
    &= 2 i \mathcal{P}_{+,a_1\cdots a_5 b_1\cdots b_5} k^{[b_1}C^{b_2 \cdots b_4]}(k)\,,
\end{align}
we reproduce the two-point function in \cref{two-point_FF} from the two-point function involving $C$ in \cref{two-point_CC}. 
In \cite{Alvarez-Gaume:1983ihn}, the propagator of an ordinary anti-symmetric field strength was used in the anomaly computation. This propagator can be obtained from \cref{two-point_FF} by simply stripping off the projectors.
In a theory of only self-dual fields, this can be done as long as we consider gravitation coupling with only the self-dual part of the field strength, i.e. we consider stress tensor of the self-dual field. Thus only the self-dual field strength can be emitted or absorbed at the three-point vertex. 
Let us now compare the stress tensor we get from our potential based formalism to that of \cite{Alvarez-Gaume:1983ihn}.

The gravitational coupling to the self-dual field can be obtained from its coupling to the flat space stress tensor
\begin{align}
    \mathcal{L}_{\text{int}} & = \frac{\kappa}{2} T^{ab} h_{ab}\,.
\end{align}
The flat space stress tensor of the self-dual field is given by
\begin{align}
     T_{ab} &= \frac{1}{4!} F_{a c_1\cdots c_4}F^{c_1\cdots c_4}{}_b\,.
\end{align}
From \cref{F_self_mom_6d}, we find the stress tensor as 
\begin{align}\label{stress_tensor_F}
     T_{ab} = - \frac{1}{3!}  \mathcal{P}_{+,a c_1\cdots c_4 b_1\cdots b_5}\mathcal{P}_{+b}{}^{c_1\cdots c_4 d_1\cdots d_5} k^{[b_1} k_{[d_1}   C^{b_2\cdots b_5]}C_{d_2\cdots d_5]}\,.
\end{align}
To complete the story, let us also write down the three-point vertex 
\begin{align}\label{3point_h_self}
     V_{h^{ab}, C^{b_1\cdots b_4} , C^{d_1\cdots d_4}} (k_1,k_2) = - \frac{\kappa}{3!} \mathcal{P}_{+,a c_1\cdots c_4 b_1\cdots b_5}\mathcal{P}_{+b}{}^{c_1\cdots c_4}{}_{ d_1\cdots d_5} k_1^{b_5} k_2^{d_5} \,.
\end{align}
Now to see the equivalence with the proposed stress tensor, let us write the bi-spinor field defined in eq. 46 of \cite{Alvarez-Gaume:1983ihn} 
\begin{align}
    \phi_{\alpha \beta} & = \frac{1}{2^{3/4}} \left(\Gamma_{a_1 \cdots a_5} \right)_{\alpha \beta} F^{a_1 \cdots a_5}\,,\cr
    & = \frac{i}{\sqrt{2}} \left(\Gamma_{a_1 \cdots a_5} \right)_{\alpha \beta} \mathcal{P}_+^{a_1 \cdots a_5 b_1 \cdots b_5} k_{[b_1}C_{b_2 \cdots b_5]}\,,
\end{align}
Here $(\alpha,\beta,\cdots)$ are spinor indices and we are using the following 10-d gamma matrix definitions
\begin{align}
    \Gamma_{a_1 \cdots a_5} = \frac{1}{5!} \gamma_{[a_1} \cdots \gamma_{a_5]}\,,\qquad \Gamma := \gamma_0 \cdots \gamma_9\,.
\end{align}
Using the following identity
\begin{align}
 \left(\Gamma_{a_1 \cdots a_5} \right)_{\alpha \beta} \epsilon^{a_1 \cdots a_5}{}_{b_1 \cdots b_5} = - 5!  \left(\Gamma_{b_1 \cdots b_5} \right)_{\alpha \delta} \Gamma_{\delta \beta}\,,
\end{align}
we can rewrite the bi-spinor field 
\begin{align}
    \phi_{\alpha \beta} = \frac{i}{\sqrt{2}} \left( \Gamma^{a_1 \cdots a_5} \right)_{\alpha \rho} k_{[a_1}C_{a_2 \cdots a_5]} \left( \mathbb{1} - \Gamma \right)_{\rho \beta}\,.
\end{align}
Thus we find
\begin{align}
    (\mathbb{1} +\Gamma)_{\alpha \alpha'} \phi_{\alpha' \beta} & = \frac{i}{\sqrt{2}} (\mathbb{1} +\Gamma)_{\alpha \alpha'}  \left( \Gamma^{a_1 \cdots a_5} \right)_{\alpha' \rho} k_{[a_1}C_{a_2 \cdots a_5]} \left( \mathbb{1} - \Gamma \right)_{\rho \beta} \cr
    & = \sqrt{2} i (\mathbb{1} +\Gamma)_{\alpha \rho} \left( \Gamma^{a_1 \cdots a_5} \right)_{ \rho \beta } k_{[a_1}C_{a_2 \cdots a_5]} \cr
    & =  \frac{\sqrt{2} i}{5!} \left( \Gamma^{b_1\cdots b_5} \right)_{\alpha \beta} \left( \delta_{[b_1}{}^{[a_1} \cdots \delta_{b_5]}{}^{a_5]} + \epsilon_{b_1\cdots b_5}{}^{a_1\cdots a_5} \right) k_{[a_1}C_{a_2 \cdots a_5]}\,,\cr
    & = \sqrt{2} i \left( \Gamma^{b_1\cdots b_5} \right)_{\alpha \beta}  \mathcal{P}_{+,b_1\cdots b_5 a_1\cdots a_5} k^{[a_1}C^{a_2 \cdots a_5]} \,.
\end{align}
We use this to rewrite the stress tensor in eq. 51 of \cite{Alvarez-Gaume:1983ihn} given in terms of the bi-spinor field as follows\footnote{We are using $X^{(a}Y^{b)}:=\frac{1}{2}(X^a Y^b + X^b Y^a)$.}
\begin{align}\label{stress_tensor_AGW}
    & T_{ab}  = \frac{1}{8} ( \mathbb{1} +\Gamma)_{\alpha \alpha'}\phi_{\alpha' \beta} ( \mathbb{1} +\Gamma)_{\gamma \gamma'} \phi_{\gamma' \delta} (\gamma_a \Gamma)_{\alpha \gamma} (\gamma_b \Gamma)_{\beta \delta} + a \leftrightarrow b \cr
    & = -\frac{1}{2} \left( \Gamma^{b_1\cdots b_5} \right)_{\alpha\beta} \left( \Gamma^{d_1\cdots d_5} \right)_{\gamma \delta} (\gamma_{(a} \Gamma)_{\alpha \gamma} (\gamma_{b)} \Gamma)_{\beta \delta} \mathcal{P}_{+\,,b_1\cdots b_5 a_1\cdots a_5} k^{[a_1}C^{a_2\cdots a_5]} \mathcal{P}_{+\,,d_1\cdots d_5 e_1\cdots e_5} k^{[e_1}C^{e_2\cdots e_5]}  \cr
    & = -\frac{1}{3!} \mathcal{P}_{+,a c_1\cdots c_4 b_1\cdots b_5}\mathcal{P}_{+b}{}^{c_1\cdots c_4 d_1\cdots d_5} k^{[b_1} k_{[d_1}   C^{b_2\cdots b_5]}C_{d_2\cdots d_5]}\,.
\end{align}
Therefore we obtain the result we found in \cref{stress_tensor_F}.

\subsubsection{One-loop hexagon}

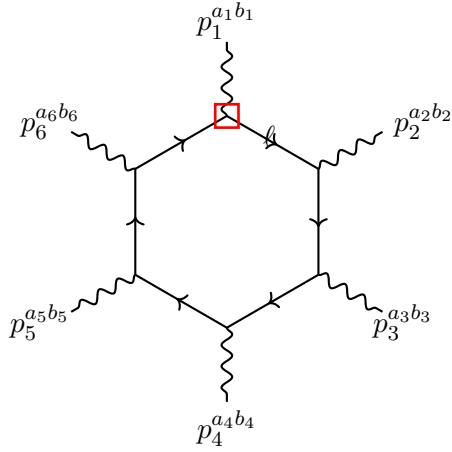
\begin{figure}[t]
\begin{center}

\begin{tikzpicture}[scale=.7, thick]


\tikzset{
    momentum/.style={
        postaction={
            decorate,
            decoration={
                markings,
                mark=at position 0.55 with {\arrow{>}}
            }
        }
    }
}


\coordinate (v1) at (90:2);
\coordinate (v2) at (30:2);
\coordinate (v3) at (-30:2);
\coordinate (v4) at (-90:2);
\coordinate (v5) at (-150:2);
\coordinate (v6) at (150:2);


\draw[momentum] (v1)--(v2);
\draw[momentum] (v2)--(v3);
\draw[momentum] (v3)--(v4);
\draw[momentum] (v4)--(v5);
\draw[momentum] (v5)--(v6);
\draw[momentum] (v6)--(v1);


\draw[decorate, decoration={snake, amplitude=2pt, segment length=8pt}]
(v1) -- ++(90:1.4);

\draw[decorate, decoration={snake, amplitude=2pt, segment length=8pt}]
(v2) -- ++(30:1.4);

\draw[decorate, decoration={snake, amplitude=2pt, segment length=8pt}]
(v3) -- ++(-30:1.4);

\draw[decorate, decoration={snake, amplitude=2pt, segment length=8pt}]
(v4) -- ++(-90:1.4);

\draw[decorate, decoration={snake, amplitude=2pt, segment length=8pt}]
(v5) -- ++(-150:1.4);

\draw[decorate, decoration={snake, amplitude=2pt, segment length=8pt}]
(v6) -- ++(150:1.4);


\node at ($(v1)+(0,1.8)$) {$p_1^{a_1 b_1}$};

\node at ($(v2)+(2.0,0.9)$) {$p_2^{a_2 b_2}$};

\node at ($(v3)+(1.6,-0.9)$) {$p_3^{a_3 b_3}$};

\node at ($(v4)+(0,-1.9)$) {$p_4^{a_4 b_4}$};

\node at ($(v5)+(-1.8,-0.9)$) {$p_5^{a_5 b_5}$};

\node at ($(v6)+(-1.6,0.9)$) {$p_6^{a_6 b_6}$};


\node at ($(v2)!0.45!(v1)+(-0.15,0.18)$) {$\ell$};


\draw[red, line width=1pt]
($(v1)+(-0.22,-0.22)$)
rectangle
($(v1)+(0.22,0.22)$);

\end{tikzpicture}
\caption{
One-loop hexagon diagram contributing to the six-graviton amplitude.
The boxed vertex denotes the insertion of the self-dual field-strength vertex operator.
}
\label{fig:hexagon}
\end{center}
\end{figure}

The one-loop hexagon diagram in \cref{fig:hexagon} can be evaluated as follows
\begin{align}
    \mathcal{A}_6 = \int \frac{d^D l}{\mathcal{D}} \mathcal{N} \,, \qquad \mathcal{D}= l^2 (l+p_2)^2 \cdots (l+p_2+\cdots +p_6)^2\,,
\end{align}
for all ingoing graviton momenta. The three-point vertex in \cref{3point_h_self} involving self-dual sector of the field strength indicated by the red box in \cref{fig:hexagon} is given by
\begin{align}
    \varepsilon^{a_1b_1}_{(1)} V_{h^{a_1b_1}, C^{i_1\cdots i_4} , C^{j_1\cdots j_4}} (k_1,k_2) & = - \frac{\kappa}{3!} \mathcal{P}_{+,a_1 c_1\cdots c_4 i_1\cdots i_5}\mathcal{P}_{+b_1}{}^{c_1\cdots c_4}{}_{ j_1\cdots j_5} k_1^{i_5} k_2^{j_5} \varepsilon^{a_1}_{(1)} \varepsilon^{b_1}_{(1)} \,,\cr
     & =: V_{IJ} (k_1,k_2)\,,
\end{align}
dressed with graviton polarization: $\varepsilon_{ab}= \varepsilon_a \varepsilon_b$. The 10-d projection operators here are defined in \cref{projection_10d}.
Here we are using compact notation $I \equiv [i_1\cdots i_4]$, etc.

For all other vertices, we shall use \textit{ordinary} graviton-matter vertex
\begin{align}
    \widetilde{V}^{I}{}_{J} (k_1,k_2) =  - \frac{\kappa}{4 \times 3!\times 5!^2} \left(\delta_{a_i}^{[i_i} \cdots \delta^{i_5]}_{c_4} \right) \left(\delta_{[j_1}^{b_i} \cdots \delta^{c_4}_{j_5]} \right) k_{1i_5} k_2^{j_5} \varepsilon^{a_i}_{(i)}\varepsilon_{(i)b_i} \,.
\end{align}

Thus the numerator can be expressed as
\begin{align}
\mathcal{N} = V^{I_1}{}_{J_1} \Pi^{J_1}{}_{I_2} \widetilde{V}^{I_2}{}_{J_2} \Pi^{J_2}{}_{I_3} \widetilde{V}^{I_3}{}_{J_3} \Pi^{J_3}{}_{I_4} \widetilde{V}^{I_4}{}_{J_4} \Pi^{J_4}{}_{I_5}   \widetilde{V}^{I_5}{}_{J_5} \Pi^{J_5}{}_{I_6} \widetilde{V}^{I_6}{}_{J_6} \Pi^{J_6}{}_{I_1}   
\end{align}
where
\begin{align}
    \Pi^{I}{}_{J} \equiv -\frac{i}{2 \times 4!} \delta^{[i_1}{}_{j_1} \cdots \delta^{i_4]}{}_{j_4}\,.
\end{align}
Let us write the numerator $\mathcal{N}$ as follows
\begin{align}
    \mathcal{N} = \mathcal{N}_{\text{even}} +  \mathcal{N}_{\text{odd}}\,, 
\end{align}
where $ \mathcal{N}_{\text{even/odd}}$ are the parity even/odd terms which are due to contractions of ($\delta-\delta$; $\epsilon-\epsilon$) and ($\delta-\epsilon$; $\delta-\epsilon$) respectively. Since we are interested in the AGW anomaly piece, we only focus on the $ \mathcal{N}_{\text{odd}}$ part. The functional dependence of this part can be expressed as
\begin{align}
     \mathcal{N}_{\text{odd}} &= \epsilon_{i_1 \cdots i_{10}} Q'^{i_1 \cdots i_{10}} (l,p_{(i)},\varepsilon_{(i)})\,,
\end{align}
due to the presence of the 10-d $\epsilon-$ tensor and Lorentz invariance. Expressing $Q'$ as
\begin{align}\label{Q_Q'_exp}
    \epsilon_{i_1 \cdots i_{10}} Q'^{i_1 \cdots i_{10}} (l,p_{(i)},\varepsilon_{(i)}) & = \epsilon_{i_1 \cdots i_{10}} Q^{i_1 \cdots i_{10}} (p_{(i)},\varepsilon_{(i)}) f\left( \ell^2,p_{(i)}\cdot p_{(j)}, p_{(i)} \cdot \varepsilon_{(j)}\right)\,,
\end{align}
the one-loop amplitude can be written as
\begin{align}
    \mathcal{A}_6 = \kappa^6 \epsilon_{i_1 \cdots i_{10}} Q^{i_1 \cdots i_{10}} (p_{(i)},\varepsilon_{(i)}) I^{(10)}\,,
\end{align}
where the 10d scalar integral $I^{(10)} := \int d^D l \frac{f(l^2)}{\mathcal{D}}$ is both UV and IR finite. The tensor structure of $Q$ can be fixed using momentum conservation and gauge invariance 
\begin{align}\label{Q_tensor}
    Q^{i_1 \cdots i_{5} j_1 \cdots j_{5}} (p_{(i)},\varepsilon_{(i)}) = \prod_{n=1}^5 p^{i_n}_{(n)} \varepsilon^{j_n}_{(n)}\,.
\end{align}
Any dependence of $\varepsilon_{(6)}$ is gauge equivalent to the above expression. For example,
consider the term $\sim p_{(i)}^4 \varepsilon_{(i)}^6$, this is equivalent to \cref{Q_tensor} due to gauge invariance: under $\varepsilon_{(i)} \to p_{(i)}$ and momentum conservation: $\sum_{i=1}^6 p_{(i)}=0$.

To see $I^{(10)}$ is both UV/IR finite, let us do a quick mass dimensional analysis. In $D$ dimensions, any $n$-point amplitude has mass dimension
\begin{align}
    [\mathcal{A}_n]_m = D - \frac{n(D-2)}{2}\,,
\end{align}
thus $[\mathcal{A}_6]_m = -14$. The gravitational coupling has mass dimensions
\begin{align}
    [\kappa]_m = -[h_{\mu \nu}]_m = -\frac{D-2}{2}\,.
\end{align}
So in $D=10$, we have $[\kappa]_m = -4$. So we have $\left[\frac{\mathcal{A}_6}{\kappa^6}\right]_m = 10$. Thus from \cref{Q_Q'_exp}, we have $[f(l^2)]_m = 7$. So we can take
\begin{align}
    f(l^2,\cdots ) \sim \left[(p_{(i)}\cdot \varepsilon_{(j)})^5 (p_i \cdot p_j), \cdots \right] \,,
\end{align}
where all other $\ell$ dependent terms can be expressed in terms of external data after tensor reduction. Now consider the scalar integral
\begin{align}
    I^{(10)} = (p_{(i)}\cdot \varepsilon_{(j)})^5 (p_i \cdot p_j) \int d^D \ell \frac{1}{\mathcal{D}}
\end{align}
and using Feynman parametrization, we write
\begin{align}
    I = \int d^D \ell \frac{1}{\mathcal{D}} = \frac{1}{5!} \prod_{i=1}^6 \int_0^1 d \alpha_i \int \frac{d^{10} \ell}{[\tilde{\ell}^2 -\Delta (\alpha_i,p_i)]^6} \propto \Gamma\left(1\right) \to \text{UV finite}\,.
\end{align}
To show IR finiteness, let us write
\begin{align}
     I = \int d^D \ell \frac{1}{\mathcal{D}} \xrightarrow{\ell \to 0} \frac{1}{ f(p_i\cdot p_j)} \int d^D \ell \frac{1}{ \ell^2 (2p_2\cdot \ell) (-2p_1\cdot \ell) }\,,
\end{align}
since the denominator takes the form $\mathcal{D}= \ell^2 (\ell+p_2)^2 \cdots (\ell-p_1)^2 \xrightarrow{\ell \to 0} \ell^2 (2p_2\cdot \ell) (-2p_1\cdot \ell) f(p_i\cdot p_j)$ and therefore in $D=10$ dimensions, $I$ is also IR finite.
This completes the demonstration of UV/IR finiteness of the one-loop integral and thus leads to anomaly. 

We conclude this section by comparing our result with that of \citep{Alvarez-Gaume:1983ihn}. The tensor structure of the one-loop amplitude in \cref{Q_tensor} coincides, up to numerical factors, with $R(\varepsilon^{(i)},p^{(j)})$ defined in equation (25) of \citep{Alvarez-Gaume:1983ihn}. Similarly, the scalar integral $I^{(10)}$ may be identified with the quantity $Z$ appearing in equation (53). Thus, our expression reproduces the structure of the anomaly obtained in \citep{Alvarez-Gaume:1983ihn}.


\subsection{On-Shell Action of Type IIB Supergravity on
\texorpdfstring{$\mathrm{AdS}_5\times S^5$}{AdS5 x S5}}
\label{sec:ads_physical_onshell}

Holography requires the renormalised type IIB supergravity action on $\mathrm{AdS}_5\times S^5$ to be non-zero \citep{Henningson:1998gx,Liu:1998bu,Gubser:1998vd,Blau:1999vz,Burgess:1999vb,Russo:2012ay}. More precisely, after reduction on $S^5$, its on-shell value must take the form \begin{equation}\label{eq:holographic_onshell_premise} S_{\mathrm{IIB}}^{\mathrm{ren}} \bigl[\mathrm{AdS}_5\times S^5\bigr] = \mathcal N_{\mathrm{hol}}\, \operatorname{Vol}_{\mathrm{reg}}\!\left(\mathrm{AdS}_5\right), \end{equation} where the coefficient $\mathcal N_{\mathrm{hol}}$ is given later in this section. This requirement is not reproduced directly by either the conventional type IIB pseudo-action or Sen's action. On the $\mathrm{AdS}_5\times S^5$ solution, the conventional ten-dimensional pseudo-action vanishes, even though the reduced five-dimensional action is proportional to $\operatorname{Vol}_{\mathrm{reg}}(\mathrm{AdS}_5)$. Authors of \citep{Kurlyand:2022vzv} resolved this mismatch by supplementing the pseudoaction with a topological boundary term whose on-shell contribution reproduces the required holographic value \citep{Kurlyand:2022vzv}. Similarly, in Sen's formulation the bulk action vanishes on-shell, and in \citep{Chakrabarti:2022jcb} it was necessary to add a pure boundary term with a specific coefficient fixed by matching to the AdS/CFT result. The two proposals agree on-shell, although their off-shell realisations are different. In the present potential-based formulation, by contrast, the action itself produces a non-zero boundary contribution upon imposing the equations of motion. We shall show that its physical part has precisely the form required by \eqref{eq:holographic_onshell_premise}, without adding an independent boundary term by hand. Holography is then needed only to fix the remaining overall normalisation of the physical five-form sector. This last point is unavoidable at the level of classical bulk dynamics. For any non-zero constant $a$, the actions $S$ and $a \, S$ for any constant $a$ 
produce identical classical equations of motion. The bulk equations and the on-shell comparison of field strengths can therefore determine the structure of the action and the relative normalisation of its terms, but not its absolute overall coefficient. That coefficient is fixed only after additional physical input is supplied, here it is the holographic dictionary.

We now evaluate the physical part of the action on the
$\mathrm{AdS}_5\times S^5$ solution of type IIB supergravity. As reviewed in \pref{sec:new_SD}, the complete action in the shifted variables decomposes as
\begin{equation}
    S
    =
    S_{\mathrm{sh}}[B]
    +
    S_{\mathrm{phys}}[C]
    +
    S_{\mathrm{bdy}}[B,C],
\end{equation}
as displayed in \cref{eq:complete_shifted_action}. Both
$S_{\mathrm{sh}}[B]$ and $S_{\mathrm{bdy}}[B,C]$ contain the propagating
shadow field. This sector is absent from the type IIB spectrum entering the
holographic dictionary. The quantity that must be compared with the
supergravity prediction is therefore the purely physical contribution
$S_{\mathrm{phys}}[C]$.

The physical action has already been written in the manifestly geometric form
\begin{equation}
    S_{\mathrm{phys}}[C]
    =
    -\frac{1}{2}
    \int_{\mathcal M}
    \d C\wedge F_{(g)},
\end{equation}
where
\begin{equation}
    F_{(g)}=F-M(F),
    \qquad
    \star_gF_{(g)}=F_{(g)},
    \qquad
    \d F_{(g)}\cong0.
\end{equation}
Introducing an overall normalisation $\kappa$ for the action, we
therefore have
\begin{equation}\label{eq:ads_physical_action}
    S_{\mathrm{phys}}[C]
    =
    -\frac{\kappa}{2}
    \int_{\mathcal M}
    \d C\wedge F_{(g)}.
\end{equation}
On shell,
\begin{equation}
    \d\left(C\wedge F_{(g)}\right)
    =
    \d C\wedge F_{(g)}
    +
    C\wedge\d F_{(g)}
    \cong
    \d C\wedge F_{(g)}.
\end{equation}
Consequently,
\begin{equation}\label{eq:ads_physical_boundary_term}
    S_{\mathrm{phys,on\text{-}shell}}
    \cong
    -\frac{\kappa}{2}
    \int_{\mathcal M}
    \d\left(C\wedge F_{(g)}\right).
\end{equation}
Thus the physical sector of the potential-based action naturally becomes a
boundary functional on shell. No additional term has been appended to the
action.

The distinction from a flux-based formulation is important. The boundary
functional in \eqref{eq:ads_physical_boundary_term} exists because the physical
gauge potential $C$ remains a fundamental variable. It is this term that can
carry a non-zero physical on-shell value even though the complete five-form
flux is self-dual.

On the $\mathrm{AdS}_5\times S^5$ background, the physical flux is identified
with the type IIB RR five-form,
\begin{equation}\label{eq:ads_flux_decomposition}
    F_{(g)}
    \cong
    F^{(5)}
    =
    F^{(5)}_{\mathrm{AdS}}
    \oplus
    F^{(5)}_{S},
\end{equation}
where
\begin{equation}
    F^{(5)}_{S}
    =
    \star_gF^{(5)}_{\mathrm{AdS}}.
\end{equation}

In an electric description, the exterior derivative of the RR four-form
potential represents its $\mathrm{AdS}_5$ component,
\begin{equation}\label{eq:ads_electric_polarisation}
    \d C^{(4)}
    \cong
    F^{(5)}_{\mathrm{AdS}},
\end{equation}
while the $S^5$ component is supplied by the self-dual completion.

Using the local on-shell potential dictionary
\begin{equation}
    C\cong C^{(4)}
\end{equation}
established in \cref{eq:RR_potential_dictionary}, we may therefore choose
\begin{equation}\label{eq:ads_C_electric}
    \d C
    \cong
    F^{(5)}_{\mathrm{AdS}}.
\end{equation}
Substituting \cref{eq:ads_flux_decomposition,eq:ads_C_electric} into
\eqref{eq:ads_physical_action} gives
\begin{align}
    S_{\mathrm{phys,on\text{-}shell}}
    &\cong
    -\frac{\kappa}{2}
    \int_{\mathcal M}
    F^{(5)}_{\mathrm{AdS}}
    \wedge
    \left(
        F^{(5)}_{\mathrm{AdS}}
        +
        F^{(5)}_{S}
    \right)
    \nonumber\\
    &=
    -\frac{\kappa}{2}
    \int_{\mathcal M}
    F^{(5)}_{\mathrm{AdS}}
    \wedge
    F^{(5)}_{S}.
\end{align}

The non-zero result therefore arises from the pairing of the electric and
magnetic parts of the self-dual flux:
\begin{equation}\label{eq:ads_electric_magnetic_pairing}
    \left.
    S_{\mathrm{on\text{-}shell}}
    \right|_{\mathrm{physical}}
    \cong
    -\frac{\kappa}{2}
    \int_{\mathcal M}
    F^{(5)}_{\mathrm{AdS}}
    \wedge
    F^{(5)}_{S}.
\end{equation}
This also explains why the result is not proportional to
$F^{(5)}\wedge F^{(5)}$, which vanishes identically for a five-form. The
potential selects one polarisation of the self-dual field, while $F_{(g)}$
contains both electric and magnetic components.

Following \citep{Chakrabarti:2022jcb}, the RR $5$-form flux solution is

    \begin{equation} \label{eq:Ads5}
F^{(5)} \cong 4 R_{AdS}^{-1}\left(\epsilon_{\mathrm{AdS}}+\star_g \epsilon_
{\mathrm{AdS}}\right) \;.
\end{equation}

We have denoted $\epsilon_{\mathrm{AdS}}$ as the epsilon tensor in $A d S_5$. $R_{AdS}$ is the radius of $\mathrm{AdS}_5$ space given by $R_{AdS}^4=4 \pi \alpha^2 g_s N$, with $\alpha^{\prime}$ is the Regge slope, $g_s$ is the string coupling, and $N$ stands for the number of quanta of RR 5 -form flux.

As pointed out in \citep{Kurlyand:2022vzv}, the contribution on-shell from the Einstein-Hilbert action, the GHY term vanish. The other fields do not contribute since on-shell they are all zero as well. Therefore, for the bosonic part of type IIB supergravity action, the only contribution to on-shell action comes from the $5$-form flux. For our potential-based formulation it is given by \cref{eq:ads_electric_magnetic_pairing}.

Using the solution given in \cref{eq:Ads5},
\begin{equation}
    F^{(5)}_{\mathrm{AdS}}
    =
    \frac{4}{R_{AdS}}\,\epsilon_{\mathrm{AdS}},
    \qquad
    F^{(5)}_{S}
    =
    \frac{4}{R_{AdS}}\,\star_g\epsilon_{\mathrm{AdS}} = \frac{4}{R_{AdS}}\,\epsilon_{\mathrm{S}},
\end{equation}
where and $\epsilon_S = \star_g \epsilon_{\mathrm{AdS}}$ is the physical volume form of the $S^5$,
we may equivalently write
\begin{equation}
    F^{(5)}_{\mathrm{AdS}}
    =
    4 R_{AdS}^4\widehat\epsilon_{\mathrm{AdS}},
    \qquad
    F^{(5)}_S
    =
    4 R_{AdS}^4\widehat\epsilon_S,
\end{equation}
in terms of unit-radius volume forms denoted by $\widehat{\epsilon}$. With the orientation
\begin{equation}
    \widehat\epsilon_{10}
    =
    \widehat\epsilon_{\mathrm{AdS}}
    \wedge
    \widehat\epsilon_S,
\end{equation}
one obtains
\begin{equation}
    \int_{\mathcal M}
    F^{(5)}_{\mathrm{AdS}}
    \wedge
    F^{(5)}_S
    =
    16\rho^8\,
    \operatorname{vol}_{\mathrm{reg}}
    \left(\mathrm{AdS}_5\right)
    \operatorname{vol}\left(S^5\right),
\end{equation}
where the volumes on the right-hand side are computed using the corresponding
unit-radius metrics. Hence
\begin{equation}\label{eq:ads_physical_onshell_result}
    \boxed{
    \left.
    S_{\mathrm{on\text{-}shell}}
    \right|_{\mathrm{physical}}
    \cong
    -8\kappa R_{AdS}^8\,
    \operatorname{vol}\left(S^5\right)
    \operatorname{vol}_{\mathrm{reg}}
    \left(\mathrm{AdS}_5\right).
    }
\end{equation}

This has precisely the form required by holography: after integrating over the
internal sphere, the physical ten-dimensional action is a fixed constant
multiplying the regularised $\mathrm{AdS}_5$ volume. Comparing
\eqref{eq:ads_physical_onshell_result} with the required value
in the convention of \citep{Kurlyand:2022vzv,Chakrabarti:2022jcb} fixes the overall normalisation of the self-dual
sector,
\begin{equation}\label{eq:ads_kappa_matching}
    \kappa
    =
    -\frac{1}{2\kappa_{10}^{2}},
\end{equation}
with $\kappa_{10}$ being the $10$-dimensional Newton's constant.

The essential point for our formalism is that the holographic comparison does not require an
additional boundary term. The boundary functional
\eqref{eq:ads_physical_boundary_term} is already implied by the
potential-based action. Holography fixes only its overall normalisation. This
differs from the flux-based treatment of \citep{Chakrabarti:2022jcb}, and from
the modification of the pseudo-action in \citep{Kurlyand:2022vzv}, where the
corresponding contribution had to be supplied separately. This does suggest that the potential-based formulation, once embedded into a spacetime action for string theory would be a natural point of comparison for holographic consideration. Investigations into a stringy origin of the new formalism is an exciting direction to which we hope to return in the future.


\section{Conclusion and Outlook} \label{sec:conclusion}

In this paper, we extended the potential-based formulation of self-dual and duality-symmetric gauge theories developed in Part I \citep{Chakrabarti:2025gyt} to curved spacetime. The construction is based on the same metric-dependent map introduced in Sen's formalism \citep{Sen:2015nph,Sen:2019qit,Andriolo:2020ykk,Hull:2023dgp}, which allows the physical self-dual field strength to be constructed directly from gauge potentials while maintaining covariance in a curved background.

A central feature of the formalism is the higher-form $h$-gauge symmetry, which relates the potential-based description to Sen's formulation, and realise both as different gauge fixing choices of the new parent formalism. One gauge choice reproduces Sen's theory, while another yields a formulation entirely in terms of ordinary gauge potentials, with the shadow sector decoupling directly at the level of the action. We also generalised the construction to an arbitrary reference metric, thereby incorporating the two-metric framework of \citep{Hull:2023dgp}. An important observation is that the $h$-gauge symmetry itself depends on the reference metric. This makes it qualitatively different from conventional gauge symmetries, whose action is independent of background geometric structures. Understanding the deeper origin and interpretation of this novel symmetry remains an important open problem.

Starting from self-dual theories in $D=4n+2$, we derived curved-space formulations of duality-symmetric theories in lower and more general dimensions. This includes manifestly electric-magnetic duality-symmetric theories obtained by toroidal reduction as well as a polyform formulation valid in arbitrary dimension. In each case we obtained explicit expressions for the corresponding gravitational map, providing a unified framework that applies to both the potential-based and Sen formulations. 

We also subjected the curved-space theory to two non-trivial physical tests. First, we showed that the formalism reproduces the expected gravitational anomalies of chiral fields, including the anomalous Ward identity of the chiral boson in two dimensions and the gravitational anomaly of the self-dual five-form in ten dimensions. These results provide a quantum check of the unconventional coupling to gravity encoded in the formalism. Second, we evaluated the on-shell action of type IIB supergravity on $\mathrm{AdS}_5\times S^5$ and found the expected non-vanishing result. Unlike earlier approaches, the required boundary contribution emerges naturally from the action itself once the gauge potential is retained as a fundamental field. Such puzzles also exist for other backgrounds. In context of PST formalism there have been some attempts \citep{Adhikari:2026rfb}. It will be interesting to do the same for our formalism as well, which unlike PST, is much more closely related to the type IIB string field theory action.

There are two immediate directions in which the present construction should be developed further. The first is to understand its stringy origin in full detail. Sen's formalism has a natural relation to string field theory, and it would be highly desirable to identify a string-field-theoretic construction whose low-energy effective action is the parent action introduced in Part I and extended here to curved spacetime. Such a derivation could explain the origin of the gauge potential, the auxiliary self-dual variable, the shadow sector, and the unusual $h$-gauge symmetry from a more fundamental perspective.

The second direction concerns global aspects of the relation between the potential $C$ and the standard RR potential $C^{(4)}$. While we established their local on-shell equivalence, a complete understanding requires addressing global issues such as large gauge transformations, flux quantisation, and the patching of higher-form gauge fields. A formulation in terms of differential cohomology or related global structures should clarify this relation and provide a deeper understanding of the boundary contribution that plays a key role in the type IIB analysis.

\section*{Acknowledgement}
 SC thanks Ashoke Sen for helpful discussions. SC thanks organisers of Indian Strings Meeting 2025, Bhubaneshwar and KIAS String Seminar series, Seoul, where initial results of this draft were presented. SC acknowledge the financial support by the Czech Science Foundation (GA\v{C}R) grant “Dualities and higher derivatives” (GA23-06498S).

\begin{appendix}

\section{Component Conventions}\label{app:component_dictionary}

The main text predominantly uses the form notation, but some sections do require components for computation. To facilitate going from one expression to the other, we summarise our conventions here.

A $p$-form $X$ is written as
\begin{equation}
    X=\frac{1}{p!}X_{\mu_1\cdots\mu_p}
    \d x^{\mu_1}\wedge\cdots\wedge\d x^{\mu_p},
\end{equation}
where its components are completely antisymmetric. For a $q$-form $Y$, the wedge product is
\begin{equation}
    X\wedge Y=\frac{1}{p!q!}
    X_{\mu_1\cdots\mu_p}Y_{\nu_1\cdots\nu_q}
    \d x^{\mu_1}\wedge\cdots\wedge\d x^{\mu_p}
    \wedge\d x^{\nu_1}\wedge\cdots\wedge\d x^{\nu_q}.
\end{equation}
We use Lorentzian signature $(-,+,\ldots,+)$ and the orientation
$\epsilon_g^{0\cdots D-1}=1/\sqrt{|\det g|}$, with indices on $\epsilon_g$ raised and lowered using $g$. The Hodge dual is
\begin{equation}
    (\star_g X)_{\mu_1\cdots\mu_{D-p}}
    =\frac{(-1)^{p(D-p)+1}}{p!}
    \epsilon_{g,\mu_1\cdots\mu_{D-p}}{}^{\nu_1\cdots\nu_p}
    X_{\nu_1\cdots\nu_p}.
\end{equation}
The reference Hodge operator $\star$ is obtained by replacing $g$ with $\eta$. For two $p$-forms, the corresponding inner product is
\begin{equation}
    \int X\wedge\star_g Y
    =\frac{1}{p!}\int\d^D x\sqrt{|\det g|}\,
    X_{\mu_1\cdots\mu_p}
    g^{\mu_1\nu_1}\cdots g^{\mu_p\nu_p}
    Y_{\nu_1\cdots\nu_p}.
\end{equation}

Once again inner-product with respect to the reference metric is obtained replacing $g$ with $\eta$.

For $p=2n+1$ in $D=4n+2$, the gravitational map has the component expression
\begin{equation}
    [M(F)]_{\mu_1\cdots\mu_p}
    =\frac{1}{p!}
    M_{\mu_1\cdots\mu_p}{}^{\nu_1\cdots\nu_p}
    F_{\nu_1\cdots\nu_p},
\end{equation}
where the kernel is antisymmetric in each group of indices and depends algebraically on the physical and reference metrics as elaborated in the next appendices.

\section{Construction of the map $M$}\label{app:construction_M}

We use the construction of \citep{Andriolo:2020ykk} for the linear map $M$ in the main text satisfying the following properties
\begin{align}
    \star_g (F-M(F)) = (F-M(F)) \,, \qquad \star M(F) = -M(F)\,,
\end{align}
where $F$ a self-dual form w.r.t the flat $\eta$. The map also satisfies the symmetry property: $F_1 \wedge M(F_2) = F_2 \wedge M(F_1)$ with $F_{1,2}=\star F_{1,2}$. In this appendix we briefly review their construction. 

Let us consider a set of basis of ($2n+1$)-forms in ($4n+2$) dimensional flat spacetime  
\begin{align}
    \lambda_{\pm}^I \,, \quad \star \lambda_{\pm}^I =\pm \lambda_{\pm}^I\,,
\end{align}
where $I$ runs from $\left[1, \cdots ,\frac{1}{2} \binom{4n+2}{2n+1} \right]$.
The action of $M$ on this basis is given by 
\begin{align}
    M(\lambda_+^I) = M^{I}{}_J \lambda_-^J \,,  \qquad  M(\lambda_-^I) =0\,.
\end{align}
In curved spacetime, we similarly introduce a basis of self-dual forms
\begin{align}
    \Lambda_{+}^I : \quad \star_g \Lambda_{+}^I = \Lambda_{+}^I\,.
\end{align}
Locally on a coordinate patch, this basis can be related to the flat space basis as  
\begin{align}\label{eq:basis_relation} 
    \Lambda^I_+ = \mathcal{S}^I{}_J \lambda_+^J + \mathcal{A}^I{}_J \lambda_-^J\,,
\end{align}
for some appropriate matrices $\mathcal{S}$ and $\mathcal{A}$. The linear map $M$ can then be expressed in terms of these matrices. 
To determine its explicit form, we consider again the specific combination of basis elements $\lambda_{\pm}^I$ that is self-dual w.r.t the physical metric $g$ 
\begin{align}
    \lambda_+^I -\mathcal{M}^{I}{}_J \lambda_-^J =\star_g \left(\lambda_+^I -\mathcal{M}^{I}{}_J \lambda_-^J \right)\,.
\end{align}
However, since $(\lambda_+^I -\mathcal{M}^{I}{}_J \lambda_-^J)$ is a self-dual form in curved spacetime, it can be expanded in terms of the basis $\Lambda_+^I$ 
\begin{align}
     \lambda_+^I -\mathcal{M}^{I}{}_J \lambda_-^J = \mathcal{P}^I{}_J \Lambda_{+}^J \,.
\end{align}
Substituting \cref{eq:basis_relation} and then comparing the coefficients of $\lambda_{\pm}$, we obtain
\begin{align}
    \mathcal{M}=-\mathcal{S}^{-1}\mathcal{A}\,.
\end{align}
This completes the construction of the linear map $M$.

\section{Dimensionally reduced $M$ using general torus metric}\label{app:4d_torus_complex_modulus}

The flat torus metric in terms of the complex structure is given by\footnote{ $A= (2\pi R)^2$ where $x^a \sim x^a + 2\pi R$.}
\begin{align}
    G_{ab} = \frac{A}{\tau_2} \begin{pmatrix}
        1 & \tau_1 \\
        \tau_1 & |\tau|^2
    \end{pmatrix}
\end{align}
where $\tau = \tau_1 +i \tau_2$. Let us call the directions along the $T^2$ directions as $\{ y^a\}$. The Hodge star action along these directions si then given by
\begin{align}
    \star_G dy^a = \Omega^a{}_b dy^b = \Omega^{ac}\epsilon_{cb} dy^b =\frac{1}{\sqrt{G}}G^{ab}\epsilon_{bc}dy^c\,,
\end{align}
where 
\begin{align}\label{omega_matrix_reps}
 \epsilon_{cb} =\begin{pmatrix}
     0 & 1\\
     -1 & 0
 \end{pmatrix}\,, \quad    \Omega^a{}_b = \frac{1}{A^2 \tau _2} \left(
\begin{array}{cc}
 \tau _1 & |\tau ^2| \\
 -1 & -\tau _1 \\
\end{array}
\right)\,.
\end{align}
To construct the linear map $M$ for 4d theory, we consider the 6d theory and compactify the $T^2$ directions. So let us consider the self and anti-self dual bases for any generic 3-form in 6d. The bases can be written as
\begin{align}
    \lambda_+^I & = e^I \wedge dy^1 + \tilde{e}^I \wedge dy^2\,,\cr
    \lambda_-^I & = e^I \wedge dy^1 - \tilde{e}^I \wedge dy^2\,,
\end{align}
where $e^I$ and $\tilde{e}^I = \star_{\eta_4} e^I$ are the bases for the $2-$forms in 4d we get after the reduction: $F= F_I e^I$ and $\widetilde{F}=\star_{\eta_4}F =\widetilde{F}_I \tilde{e}^I$ with $I \in[0,\cdots,3]$. Here we used $\star_{\delta_2} dy^1 =\star_{\delta_2} dy_1 =\epsilon_{12} dy^2 = dy^2$ and $\star_{\delta_2} dy^2 =- dy^1$, $\delta_2$ being the flat 2-d metric. Now let us consider a self-dual basis $\Lambda_+^I$ such that
\begin{align}
    \Lambda_+^{Ia} = \star_{g_6} \Lambda_+^{Ia} = (\star_{g_4} \otimes \star_{G}) \Lambda_+^{Ia} \,,
\end{align}
where we suppress the $T^2$ indices below.
Since this is a rank-3 form in 6d, we can write
\begin{align} \label{int_calc_1}
    \Lambda_+^{I} & = \mathcal{S} \lambda_+^I + \mathcal{A} \lambda_-^I\cr
    & = (\mathcal{S} + \mathcal{A}) e^I \wedge dy^1 + (\mathcal{S} - \mathcal{A}) \tilde{e}^I \wedge dy^2 \,,
\end{align}
and the linear map $M$ can be expressed as $M=-\mathcal{S}^{-1}\mathcal{A}$. If we focus on the $2-$form sector after reduction of $\Lambda_+^{I}$, we may write
\begin{align}\label{int_calc_2}
     \Lambda_+^{I} = e^I \wedge dy^1 + \star_{g_6}(e^I \wedge dy^1)\,,
\end{align}
where using the representation of $\Omega$ matrix in \eqref{omega_matrix_reps}
\begin{align}
    \star_{g_6}(e^I \wedge dy^1)= (\star_{g_4} e^I \wedge \star_{G} dy^1) & = E^I \wedge \star_{G} dy^1 \cr
    & = \frac{1}{A^2\tau_2} E^I \wedge  \left(\tau_1 dy^1 + |\tau|^2 dy^2 \right)\,.
\end{align}
By comparing \eqref{int_calc_1} and \eqref{int_calc_2}, we get
\begin{align}
    \mathcal{S} + \mathcal{A} &= \mathbb{1} + \frac{\tau_1}{A^2 \tau_2} \star_{g_4} \cr
    \mathcal{S} - \mathcal{A} &= -\frac{|\tau|^2}{A^2 \tau_2} \star_{g_4} \star_{\eta_4}\,.
\end{align}
For square metric case, we set $\tau_1 =0$ and $\tau_2 =1$ i.e. $|\tau|^2=1$, we get 
\begin{align}
    \mathcal{S} + \mathcal{A} &= \mathbb{1} \cr
    \mathcal{S} - \mathcal{A} &= -\frac{1}{A^2 } \star_{g_4} \star_{\eta_4}\,.
\end{align}

\end{appendix}

\bibliography{refs}
\end{document}